\documentclass{ws-ijmpd}
\usepackage{amssymb}
\usepackage{amsmath}
\usepackage{amsfonts}
\usepackage{hyperref}
\hypersetup{colorlinks,urlcolor=black,citecolor=black,linkcolor=black,filecolor=black}
\usepackage[super,compress]{cite}
\usepackage{graphicx}
\usepackage{float}

\begin{document}

\markboth{Leila Maghlaoui, Peter O. Hess}
{Consequences of a minimal length}

\title{Consequences of a minimal length in a pseudo-complex
extension of General Relativity
}

\author{Leila Maghlaoui}

\address{Physics Deptartment, University of Mentouri,
Constantine 1, Constantine  P.O. box,
325 Ain El Bey Way, 25017 Constnantine, Algeria \\
and\\
Mathematical and Subatomic Physics Laboratory (LPMPS),
University of Mentouri, Constantine 1, Algeria}

\author{Peter O. Hess}

\address{Instituto de Ciencias Nucleares, Universidad Nacional
Aut\'onoma de M\'exico, Ciudad Universitaria,
Circuito Exterior S/N, A.P. 70-543, 04510 M\'exico D.F. Mexico.\\
and\\
Frankfurt Institute for Advanced Studies, J. W. von Goethe University, Hessen, Germany,\\ hess@nucleares.unam.mx
}

\maketitle

\begin{history}
\received{Day Month Year}
\revised{Day Month Year}
%\accepted{Day Month Year}
%\comby{(xxxxxxxxxx)}
\end{history}

\begin{abstract}
The effects of a minimal length are investigated within
an algebraically extended theory of General Relativity (GR).
Former attempts, to include a minimal length in GR 
are first resumed, with a conformal factor of the metric as a consequence. 
Effective potentials for various black hole masses
(as ratios to the minimal length) are deduced.
It is found that the existence of a minimal length has,
for a small mass black hole,
important effects on the effective potential near the
%\color{red}
event horizon, creating barriers which inhibit that 
particles can pass the event horizon. Further, a 
%\color{black}
new limit for the minimal mass of a black hole is derived.
\end{abstract}

\ccode{PACS numbers:}

%\tableofcontents

\section{Introduction}

General Relativity (GR) is a well established theory 
for describing numerous observations in the solar system
\cite{solar} and in cosmology, 
it is one of the best tested
theories.

Nevertheless, it is not decided if GR is complete
to describe phenomena in very strong gravitational fields,
as in the vicinity of a black hole. 
For example, quantization effects are
expected to play a role. There are further problems, as the
loss of information and the appearance of an event horizon
which disconnects an outer observer to the interior of
a black hole 
%\color{red}
(though, many do not think that this is a problem).
%\color{black}

Therefore, in very
strong gravitational fields one can speculate on how 
possibly to extend GR.
Einstein himself intended to modify his theory such that it
contains Electrodynamics \cite{einstein1,einstein2}. In order to do so, he introduced a complex metric.
Another approach was followed by M. Born \cite{born1,born2},
who complained about the dominant role of coordinates in GR,
contrary to Quantum Mechanics where the momentum is
treated on an equal footing. M. Born 
tried to solve the problem by
introducing the concept of 
{\it complementarity}, modifying the length element such that
it is defined in the coordinate-momentum space. 

Later, Caianiello and his coworkers retook the ideas
of M. Born and wrote a
series of papers about a theory called Quantum Geometry 
\cite{3,4,5,6,7,8,9}
%\color{red}
(for the formulation in differential geometry see \cite{21}). 
%\color{black}
In \cite{26} a pseudo-complex 
extension of GR (pcGR) was proposed, with many applications 
\cite{book}, and in \cite{PPNP} the most recent review 
is given including several predictions. M. Born's theory
corresponds to a particular limit in pcGR \cite{PPNP}.
It was shown
that deviations to GR, for macroscopic black holes,
 only appear near the event horizon
%\color{red}
(not having yet being resolved due to resolution problems).
%\color{black}
In all applications of pcGR the minimal length was neglected,
because its effects where considered to be negligible.

In earlier contributions \cite{feoli1999} (and references
therein) the effects of a minimal length on the structure
near the event horizon were investigated within GR.
The questions, we would like to answer, are: What happens
when GR is extended algebraically to pcGR, which includes
the effects of a dark energy in the vicinity of a 
black hole? In what range of the black hole mass
the effects of a minimal length are noticeable?
In \cite{feoli1999} the effects of the {\it maximal
acceleration}, as a parameter, were studied and not
the role of a minimal length and its relation to the mass
of the black hole (though, they are related).
As we will see the minimal length affects small mass 
black holes, which may have been created during the
Big Bang.

The inclusion of a minimal
length in the extended theory 
is quite involved and we have to recur to 
approximations, as restricting to the Schwarzschild
solution only.

The paper is structured as follows: In Section \ref{section2}
the theory proposed by E. R. Caianiello is resumed. In
Section \ref{section3} 
the effects of the minimal length in pcGR 
in the Schwarzschild case is discussed and finally in Section
\ref{section4} the results are interpreted and conclusions
are drawn.

\section{The theory proposed by E. R. Caianiello}
\label{section2}

This model \cite{3,4,5,6,7,8,9,feoli1999} interprets 
the quantization via a curvature of the relativistic
eight-dimensional space-time tangent bundle 
$TM=M_{4}+TM_{4}$, with $M_{4}$ as the
usual flat space-time manifold with the metric  
$\eta_{\mu\nu}$ (see also \cite{21}). 
This description satisfies the Born reciprocity principle
\cite{born1,born2} and it incorporates the notion
that the proper accelerations of massive particles along their worldliness with upper limit
$A_{m}$, referred to as maximal
acceleration (MA).
Indeed the usual Minkowski line element 
$d^2s=\eta_{\mu\nu}dx^{\mu}dx^{\nu}$ 
is replaced by 
the line element in the eight-dimensional space-time tangent bundle TM, with $\eta_{\mu\nu}$ substituted by $g_{\mu\nu}$, 

\begin{equation}
d^{2}w=g_{AB}dX^{A}dX^{B},\text{ \ }A,B=0,....7,\text{\ }  \label{1} 
\end{equation}
where

\begin{align}
g_{AB} & =g_{\mu\nu}\otimes g_{\mu\nu}, \nonumber \\
X^{A} & =\left( x^{\mu},\frac{c^{2}}{A_{m}}\frac{dx^{\mu}}{ds}\right) ,\text{
\ }\mu=0,..3
~~~,
\end{align}
$x^{\mu}=(ct,\overrightarrow{x})$ is the usual space-time four-vector and $%
\overset{\cdot}{x}^{\mu}=u_{\mu}=\frac{dx^{\mu}}{ds}$ the four-velocity, $c$
is the speed of light in the vacuum. The velocity
components are 
strictly speaking only valid in flat space \cite{PPNP}
and it is a consequence of the dispersion relation. 
As noted above,
E.R. Caianiello substituted the $\eta_{\mu\nu}$ by a
general metric $g_{\mu\nu}$, where the relation is approximate. 

The line element in the ordinary four-dimensional space-time $%
\mathbf{M}_{4}$, in which a particles moves when the constraints of a
maximal acceleration is present, is
rewritten such that the length element has the
usual form  augmented by a conformal factor. 
In fact, we can calculate the effective
four dimensional metric $\widetilde{g}_{\mu\nu}$ on the hypersurface
locally embedded in $\mathbf{TM}_{4}$. In this case, the line element
becomes 

\begin{align}
d^{2}w & =(1+\frac{g_{\mu\nu}a^{\mu}a^{\nu}}{A_{m}^{2}})d^{2}s \label{2}  \\
d^{2}w & =\widetilde{g}_{\mu\nu}dx^{\mu}dx^{\nu} \label{3} 
\\
\text{with }d^{2}s & =g_{\mu\nu}dx^{\mu}dx^{\nu}\text{ and }\widetilde {g}%
_{\mu\nu}=(1+\frac{g_{\mu\nu}a^{\mu}a^{\nu}}{A_{m}^{2}})g_{\mu\nu}=%
\sigma^{2}(r)g_{\mu\nu}\label{4}
~~~,
\end{align} 
with the four-acceleration 
$\overset{\cdot\cdot}{x}^{\mu}=a^{\mu}=
c^2\frac{d^{2}x^{\mu}}{d^{2}s}$
($ds=cd\tau$, with $\tau$ as the eigentime). 

It is quite involved 
to obtain the exact form for the conformal factor 
$\sigma^2 (r)$ and
one recurs to an {\it iterative procedure}: 
In the first step, the
equations of motion with $\sigma^2(r)=1$ are deduced in the 
usual manner, solving the geodesic equations. 
These give the components of the 4-acceleration $a^\mu_{(0)}$.
Substituting these components into $\sigma^2(r)$ results in

\begin{equation}
\sigma_{(1)}^{2}(r)=1+\frac{g_{\mu\nu}a_{(0)}^{\mu}a_{(0)}^{\nu}}{A_{m}^{2}%  
} \label{5}
\end{equation}

The index $(i)$ denotes the order of the iteration. 
We can iterate this procedure by resolving the equations of
motion with the $\sigma^{2}_{(1)}$, obtaining the next iteration 
$a^\mu_{(1)}$ for the 4-acceleration.
As a result we obtain the next iterative expression for
$\sigma^2 = \sigma^2_{(2)}$, namely

\begin{equation}
\sigma_{(2)}^{2}(r)=1+\frac{g_{\mu\nu}a_{(1)}^{\mu}a_{(1)}^{\nu}}{A_{m}^{2}
}
~~~.
\end{equation}

This iteration procedure can be continued, however, 
in general it is
stop after the second iteration and study the effects on the
solutions, the form of the effective potential, etc.

After a second iteration, the modified length
element acquires the form

\begin{align}
d^{2}w & =\left( 1+\frac{g_{\mu\nu}a_{(1)}^{\mu}a_{(1)}^{\nu}}{A_{m}^{2}}%
\right) d^{2}s \label{6} \\
\text{ }\widetilde{g}_{\mu\nu} & =\left( 1+\frac{g_{\mu\nu}a_{(1)}^{\mu
}a_{(1)}^{\nu}}{A_{m}^{2}}\right) g_{\mu\nu}\label{7}
\end{align}

Limiting values for the accelerations were also derived by several authors
on different grounds and applied to many branches of physics such as string
theory, cosmology, quantum field theory, black hole physics etc \cite{caia1999,10,11,12,13,14,15,16,17,18,19,20,21}. 
In \cite{feoli1999} this maximal acceleration is obtained as
$A_m= \frac{2mc^3}{\hbar}$, where the causal limit
was taken 
%\color{red}
into 
%\color{black}
account.
Another group starts from an eight
dimensional space by introducing pseudo-complex variables and projecting
this space to a four dimensional physical space 
\cite{22,23,24,25}. 

Up to here, the notion of a maximal acceleration is restricted
to GR, formulated with the coordinates $x^\mu$. The central
question of our contribution is: What happens when not
only a maximal acceleration is introduced, but also
the coordinates are algebraically extended \cite{kelly}
to pseudo-complex coordinates

\begin{equation}
X^\mu = x^\mu + I y^\mu
~~~,
\label{X}
\end{equation}
with $I^2=1$ \cite{kelly}.

In \cite{26} just such an 
extension of the general relativity was proposed,
called the {\it pseudo-complex General Relativity}
(pcGR), with the most
recent review given in \cite{PPNP}.
The pseudo-complex extension is proven to be the only
consistent algebraic extension of GR \cite{kelly},
not having neither ghost nor tachyon solutions.
In pseudo-complex General Relativity the space-time
coordinates are of the form
$X_{\mu }=x_{\mu }+I\frac{l}{c}u_{\mu }$,
with its pseudo-real part $x^\mu$ and its
pseudo-imaginary component, chosen to have the form 
$y^\mu=\frac{l}{c}u_{\mu }$ in analogy to \cite{caia1999}.
The component $\frac{l}{c}u_{\mu }$
is an approximation, strictly speaking valid only in
flat space, and can be associated to the components
of the tangent vector (four velocity vector) at a given space-time point.
The factor $l$ has the unit of a length, which is introduced due to
dimensional reasons. The consequence of that is the appearance of a minimal
length scale $l$ within the theory which implies a maximal acceleration. As
mentioned above. we assume 
that the minimal length scale is of order of the
Planck length ($l_{P}=\sqrt{\frac{\hbar k}{c^{3}}=}1.616199\times 10^{-35}{\rm m}$,
$k$ is the gravitational constant and $\hbar $ is the reduced Planck
constant). 

The line element in pcGR is given by

\begin{equation}
d^{2}w=g_{\mu \nu }(X,P)DX^{\mu }DX^{\nu }
~~~,
\end{equation}
where $X^\mu$ and $P_\nu$ are pseudo-complex.

Further assumptions to the metric are applied, namely that
$g_{\mu\nu}(X)=g_{\mu\nu}(x)$, i.e. that it does not depend on
the four-velocity, and with the constraint 
that the length element itself is
pseudo-real, which is the case when the dispersion relation
is satisfied \cite{PPNP}.
With this, the length element in pcGR acquires the form

\begin{equation}
d^{2}w=g_{\mu \nu }(x)\left[ dx^{\mu }dx^{\nu }+\left( \frac{l}{c}\right)
^{2}du^{\mu }du^{\nu }\right]  \label{8}
\end{equation}

In this extension the geometry is understood as a consequence of curvature
in eight-dimensional phase space, in which the coordinates of the
velocity-space manifold are the components of the 
four-velocity with a dimensional factor $l$ of
the order of the Planck length $l_{P}$. 
In (\ref{8}) the length element $ds^2=g_{\mu\nu}dx^\mu dx^\nu$
is extracted and we use $ds=cd\tau$, with $d\tau$ as the
eigentime. We obtain

\begin{align}
d^{2}w & =d^{2}s\left( 1-\left( \frac{l^2}{c^4}\right)
\left\vert
g_{\mu\nu}a^{\mu}a^{\nu}\right\vert \right)  \label{9} 
\\
d^{2}w & =\widetilde{g}_{\mu\nu}dx^{\mu}dx^{\nu} 
\nonumber \\
\widetilde{g}_{\mu\nu} & =
\left( 1-\left( \frac{l^2}{c^4}\right)
\left\vert g_{\mu\nu}a^{\mu}a^{\nu}\right\vert \right) g_{\mu\nu}
~~~,
\end{align}
with $\widetilde{g}_{\mu\nu}$ representing the modified metric
and $a^\mu$ the acceleration component $\frac{du^\mu}{d\tau}$,
with $\tau$ as the eigentime.

Therefore, one can rewrite the length element as

\begin{eqnarray}
d^{2}w & = & \sigma^{2}(r)d^{2}s  
\nonumber \\
\text{with }&&
\nonumber \\
\sigma^{2}(r) & = &
\left( 1-\left( \frac{l^2}{c^4}\right)
\left\vert g_{\mu\nu}a^{\mu}a^{\nu}\right\vert \right)
~~~.
\label{12}
\end{eqnarray}

It is clear that the maximal acceleration is included automatically
in pseudo-complex general relativity 
($A_{m}=\frac{c^2}{l}$). The
new metric $\widetilde{g}_{\mu\nu}$ depends
on the coordinates $x_{\mu}$ and on the
acceleration field $a_{\mu}$. In this case, the generalized proper-time
interval (\ref{9}) becomes 

\begin{equation}
d^{2}w=\left( 1-\frac{l^{2}}{c^{4}}\left\vert a\right\vert ^{2}\right) d^{2}s
~~~,
\label{d2w}
\end{equation}
which restricts the acceleration, as before, to a finite
interval with a  maximal acceleration 
$\frac{c^2}{l}$ = $\frac{2mc^3}{\hbar}$
\cite{feoli1999,3,4,5,6,7,8,9}. 
In \cite{27} it was applied
to a pc-field theory with the important 
feature that it is by construction
regularized, due to the appearance of a minimal length 
scale $l$. Due to the
fact that this minimal length scale is a 
{\it parameter}, it is not affected by a
Lorentz transformation and, thus, all symmetries are preserved,
which is a huge simplification compared to theories which
require a violation of the Lorentz symmetry. 
In \cite{PPNP,28}
this fact was used in the hope that it prevents the formation of a black
hole. Indeed the formation of a black hole is avoided in this theory, but
not due to the minimal length scale but due to the appearance of dark
energy which appears naturally within this theory. The same effects appear
when it was applied to the Robertson-Walker universe
\cite{book,28}. In all cases, the
variational principle introduced contributions, which can be
interpreted as dark energy, acting repulsively such that the formation of an
event horizon and a singularity at the center is avoided. This is a most
important result, as any proper theory should not contain singularities.

Finally, one has to note that the second term in 
the conformal factor $\sigma^2$ is 
only a covariant quantity in flat space, where the 
dispersion relation $\eta_{\mu\nu} dx^\mu dy^\mu = 0$,
with $y^\mu = \left(\frac{l}{c}\right)u^\mu$ 
($u^\mu$ is the 4-velocity) is valid \cite{feoli1999}. 
In a curved space,
one has to solve a complicated equation, which is the result
of the constraint that the pseudo-complex length element is
real, i.e., the pseudo-imaginary term has to vanish.
For more details and first attempts to solve it, please
consult \cite{PPNP}. This is a price to pay when the 
iteration procedure is used.

\section{The Pseudo-complex Schwarzschild geometry}
\label{section3}

In pcGR it is assumed that a central mass generates,
due to quantum effects, a
distribution of dark energy in its 
%\color{red}
vicinity, which is based on theoretical grounds using
semi-classical Quantum Mechanics \cite{visser}. 
%\color{black}
The distribution
is parametrized as $B_n/r^n$, with 
two phenomenological parameters, 
$B_n$ and $n$. The $B_n=bm^{n}$ describes the coupling of the
central mass to the dark energy and $n$ its fall-off as
a function of the radial distance.
This is the simplest ansatz and
one easily can add further complicated dependencies in $r$.
In \cite{26} $n$ was set to 2, which results in a metric already excluded by
solar system experiments \cite{solar}. In 
\cite{26 , Schon2013 , Schon2014} the $n$ was equal to $3$ and
several observable predictions were made. Using the first observation of
gravitational waves \cite{wave2016}, in \cite{Nielsen2018, Nielsen2019} it is shown that $n=3$ is also 
%\color{red}
excluded, thus $n=4$ is assumed now.
%Due to this reason, we will take $n=4$. 
%\color{black}
In this case the pseudo-complex Schwarzschild solution is 

\begin{equation}
d^{2}s=\left( 1-\frac{2m}{r}+\frac{\Omega }{2r}\right) d^{2}t-\frac{d^{2}r}{%
\left( 1-\frac{2m}{r}+\frac{\Omega }{2r}\right) }-r^{2}(d^{2}\theta +\sin
^{2}\theta d^{2}\phi )\label{13} 
\end{equation}%
\begin{equation}
\Omega (r)=\frac{B_4}{3}r^{-3}\label{14}
\end{equation}
The results for $n=3$ are quite similar, changing 
only details.

In \cite{universe} the $B_4$ parameter was varied 
within the pc-Kerr solution from zero
(GR) to a maximal value, from which on no event horizon exists
anymore. The transition from the existence of an event horizon
to its disappearance could be related to a phase transition.
However, again the effects of a minimal length were
neglected. Here, we also will consider the whole range of
$B_4$, i.e, investigating the transition from
GR to pcGR, but now with the
inclusion of a minimal length.

With the approximation applied, 
the line element with a minimal length $l$\ is given by the
relation (\ref{d2w}),
which is similar in form as given by E. R. Caianiello.

In the following sub-section the  factor 
$\sigma^{2}(l)$ will be investigated further. 

\subsection{The factor $\protect\sigma ^{2}(r)$}

In order to calculate the corrections to the pc-Schwarzschild
metric, experienced by a particle and
its motion along a geodesic, 
one must determine the
factor $\sigma ^{2}(r)$ of the order of $l^{2}$. Before that, we will impose
some conditions on the factor $\sigma ^{2}(r).$ By 
using the embedding
procedure mentioned above of the order $l^{2}$, the line element acquires the form (\ref{d2w}),
which has to be positive definite, i.e.,

\begin{equation}
d^{2}w>0\Longleftrightarrow\left\vert 
a^{2}\right\vert <\frac{c^{4}}{l^{2}}
~~~,
\label{16}
\end{equation}

This implies that the acceleration 
is limited from above by
$A_m=c^{2}/l$, called the maximal acceleration. 
With this condition,
the factor $\sigma ^{2}(r)$ is limited by 0 from below,
when the acceleration is maximal, and by 1 from above,
when the acceleration is zero:

\begin{equation}
0 \leq \sigma ^{2}(r) \leq 1\label{17}
\end{equation}
As a consequence, the pcGR implies a
maximal 
%\color{red}
acceleration, as in GR
%\color{black}
\cite{3,4,5,6,7,8,9}. 

In order to deduce the explicit 
expression of $\sigma ^{2}(r)$ we recur again to an
iteration procedure 
%\color{red}
(restricting now to the first iteration),
5\color{black} 
which provides an 
approximate solution for this factor:
In the first 
%\color{red}
iteration, to which we will restrict, 
%\color{black}
the metric with $\sigma^2=1$
is taken 
and the corresponding geodesic equations are
derived and solved. To obtain these geodesic equations,
we start from the variational integral
principle

\begin{equation}
\delta \int \left[ \left( 1-\frac{2m}{r}+\frac{B_4}{6r^{4}}\right) \left( 
\overset{\cdot }{t}_{0}\right) ^{2}-\frac{1}{\left( 1-\frac{2m}{r}+\frac{B_4}{%
6r^{4}}\right) }\left( \overset{\cdot }{r}\right) ^{2}-r^{2}(\left( \overset{%
\cdot }{\theta }\right) ^{2}+\sin ^{2}\theta \left( \overset{\cdot }{\phi }%
\right) ^{2})\right]ds^2 ~ = ~ 0
~~~,
\end{equation}
where the dot refers to the derivative with respect 
to the variable s.
The variation leads
to equations of motion

\begin{align}
\frac{d}{ds}(r^{2}\overset{\cdot }{\theta }^{2})& =r^{2}\sin \theta \cos
\theta \overset{\cdot }{\phi }^{2} \label{18} 
\\
\frac{d}{ds}(r^{2}\sin ^{2}\theta \overset{\cdot }{\phi })& =0 \label{19} 
\\
\frac{d}{ds}\left[ \left( 1-\frac{2m}{r}+\frac{B_4}{6r^{4}}\right) \overset{%
\cdot }{t}\right] & =0
~~~,
\label{20} 
\end{align}
from where the accelerations $a_{(0)}^\mu$ in this first 
iteration are obtained.

In this completely relativistic metric of the 
pc-Schwarzschild, all orbits will be still in the 
orbital plane defined by

\begin{equation}
\theta =\frac{\pi }{2};\text{ }\overset{\cdot }{\theta }=0
~~~.
\end{equation}

With this, the equations of motion for the energy and the angular
momentum, respectively, are

\begin{align}
\left( 1-\frac{2m}{r}+\frac{B_4}{6r^{4}}\right) \overset{\cdot }{t}&=E 
 \label{21} \\
r^{2}\overset{\cdot }{\phi }& =L   \label{22}
~~~.
\end{align}

In addition, we can use $\frac{ds^2}{ds^2}=1$, which leads to
a third equation, namely

\begin{equation}
1=\left( 1-\frac{2m}{r}+\frac{B_4}{6r^{4}}\right) c^{2}\overset{\cdot }{t}^{2}-%
\frac{1}{\left( 1-\frac{2m}{r}+\frac{B_4}{6r^{4}}\right) }\overset{\cdot }{r}%
^{2}-r^{2}\overset{\cdot }{\phi }^{2}  \label{23} 
~~~.
\end{equation}

Substituting Eqs. (\ref{21}) and (\ref{22}) 
into (\ref{23}), we obtain

\begin{equation}
\overset{\cdot }{r}^{2}=E^{2}-\left( 1-\frac{2m}{r}+\frac{B_4}{6r^{4}}\right)
\left( \frac{L^{2}}{r^{2}}+1\right)  \label{24} 
~~~.
\end{equation}

Substituting the last results into $\sigma^2$ =
$\left( 1-\frac{l^{2}}{c^{4}}\left\vert a^{2}\right\vert \right)$
leads to $\sigma_{(1)}^2$ in terms of
$l$, $\overset{\cdot
\cdot }{t}$, $\overset{\cdot \cdot }{r},$ and $\overset{\cdot \cdot }{\phi }$

\begin{align}
\sigma_{(1)}^{2}(r)& = 
& \left\{ 1-\frac{l^{2}}{c^{4}}\left\vert -c^{2}\left( 1-\frac{2m}{r}+\frac{B_4}{6r^{4}}\right) \overset{\cdot \cdot }{t}^{2}+\frac{1}{\left( 1-\frac{2m}{r}%
+\frac{B_4}{6r^{4}}\right) }\overset{\cdot \cdot }{r}^{2}+r^{2}\overset{\cdot
\cdot }{\phi }^{2}\right\vert \right\}\label{25} 
\end{align}

with
\begin{align}
\overset{\cdot \cdot }{t}& =\frac{-E}{c^{2}\left( 1-\frac{2m}{r}+\frac{B_4}{%
6r^{4}}\right) ^{2}}\left( \frac{2m}{r^{2}}-\frac{2B_4}{3r^{5}}\right) \overset%
{\cdot }{r} \nonumber \\
\overset{\cdot \cdot }{r}& =\left( -\frac{m}{r^{2}}+\frac{L^{2}}{r^{3}}-%
\frac{3mL^{2}}{r^{4}}+\frac{B_4}{3r^{5}}+\frac{2B_4L^{2}}{3r^{7}}\right) \nonumber \\
\overset{\cdot \cdot }{\phi }& =-\frac{2L}{r^{3}}\overset{\cdot }{r} \nonumber \\
\text{ }\overset{\cdot }{r}^{2}& =\left[ \frac{E^{2}}{c^{2}}-\left( 1-\frac{%
2m}{r}+\frac{B_4}{6r^{4}}\right) \left( \frac{L^{2}}{r^{2}}+1\right) \right]
\label{25a}
\end{align}

The $m$
is the mass of the source in units of length, 
$L$ the angular momentum 
and $E$ the energy, with $E$ and $L$ being
constant of motions, in units of the {\it particle mass} $M$.
In what follows,
we will present some plots of the factor $\sigma ^{2}(r)$, 
in order to show how 
$\sigma^{2}$ varies as a function of the
a-dimensional variables 

\begin{eqnarray}
\rho & = & 
\frac{r}{m} ~,~ \lambda ~=~ \frac{L}{m}
~,~ \epsilon ~= ~ \frac{l}{m}~,~ \alpha ~=~
\frac{B}{m^{4}}  
~~~.
\label{dimensionless}
\end{eqnarray} 

\begin{figure}[H]
\centering
\includegraphics[width=0.60\textwidth]{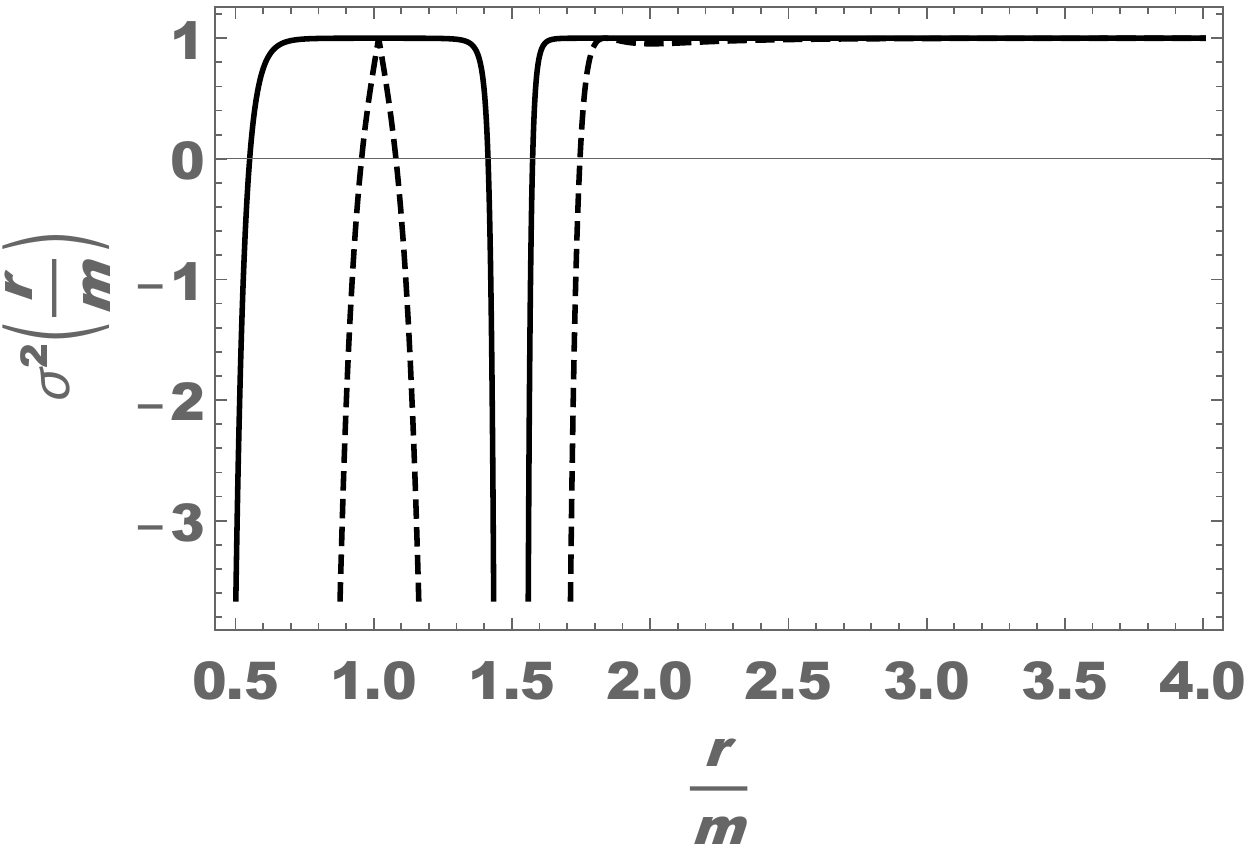} 
\caption{Dashed line: The factor $\sigma^{2}(\rho)$ 
for $\lambda=0,\alpha=81/8,\epsilon=0.1$ and $E=1$.
Solid line: The factor $\sigma^{2}(\rho)$ 
for $\lambda=0,\alpha=81/8,\epsilon=0.001$ and $E=1$.
}
\label{fig1}
\end{figure}

\begin{figure}[H]
\centering
\includegraphics[width=0.60\textwidth]{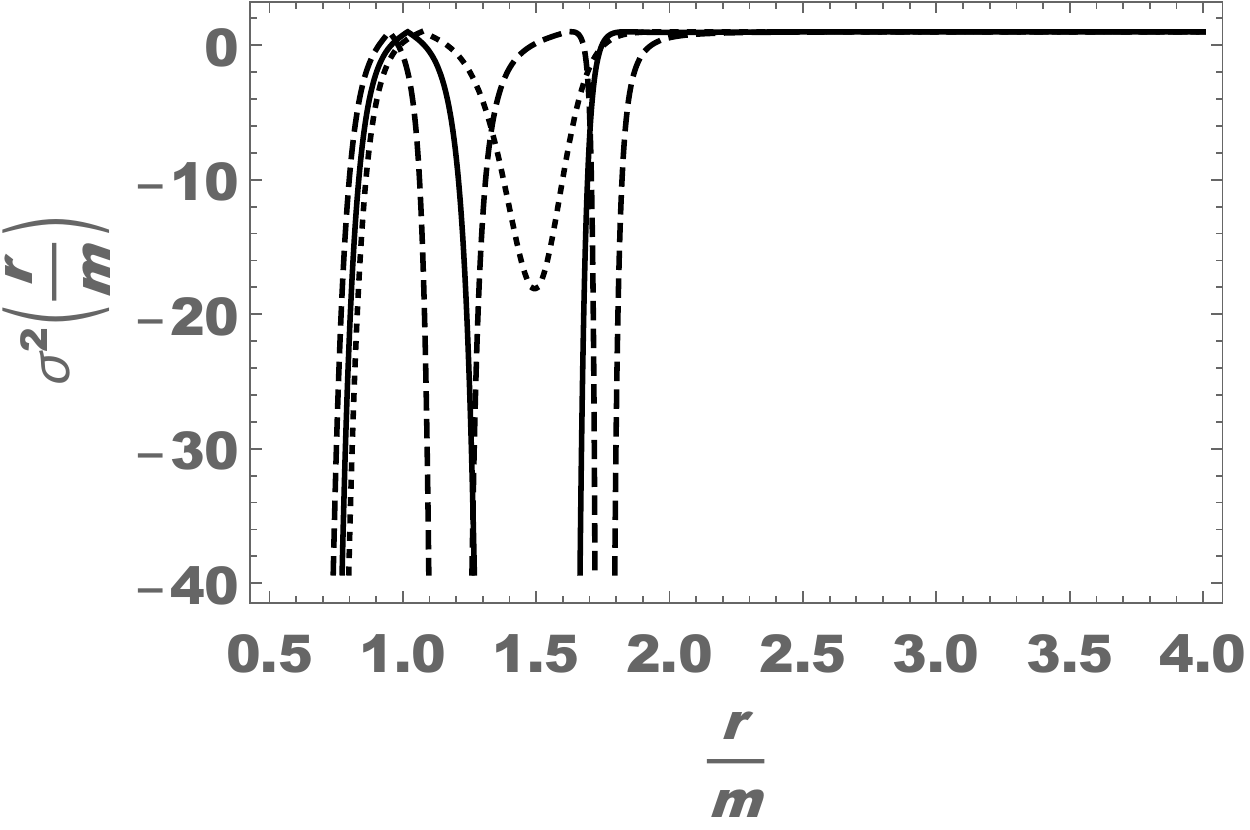} 
\caption{Dashed line: the factor
$\sigma^{2}(\rho)$ for $\lambda=0$, $\alpha=8,\epsilon=0.1$ ,
$E=1$. Solid line: the factor $\sigma^{2}(\rho)$ 
for $\lambda=0,\alpha=81/8,\epsilon=0.1$ and $E=1$ and Dotted line: the factor $\sigma^{2}(\rho)$ 
for $\lambda=0,\alpha=12,\epsilon=0.1$ and $E=1$.
}
\label{fig2}
\end{figure}

The dependence of $\sigma^2$ on different parameter values
$\epsilon$ ($l$) and $B$ ($\alpha$)
is shown in  Figs. \ref{fig1} and \ref{fig2}. As can be seen,
there are two effects: i) The deviations to $\sigma^2=1$ 
become noticeable
the more $\alpha$ approaches from above $\frac{81}{8}$,
where still an event horizon exists, for larger $\alpha$ the
deviations are smoothed out. When $\alpha$ is smaller than
$\frac{81}{8}$ the deviations increase. In all cases,
$\sigma^2$ is strongly lowered below $\frac{r}{m}=2$.
For large $\frac{r}{m}$ the $\sigma^2$ approaches 1.
Note, that $\epsilon$ = $\frac{l}{m}$ = 0.001 
implies a small mass 
only thousand time larger than the minimal length.
For a large mass, for example of the size of a 
%\color{red}
regular star, 
only small or no effects are seen.
%\color{black}

%\begin{figure}[H]
%\centering
%\includegraphics[width=0.60\textwidth]{sigma3.pdf} 
%\caption{Solid line: the factor
%$\sigma^{2}(\rho)$ for $\lambda=90$, $\alpha=8,\epsilon=0.01$ ,
%$E=10$. Dashed line: the factor $\sigma^{2}(\rho)$ for $\lambda=90,\alpha=8,\epsilon=0.1$ and $E=10$ and Dotted line: the factor $\sigma^{2}(\rho)$ for $\lambda=90,\alpha=8,\epsilon=0.00001$ and $E=10$.}
%\label{fig3}
%\end{figure}

To resume, $\sigma ^{2}(\rho =\frac{r}{m})\longrightarrow 1$ 
as $\rho$ = $\frac{r}{m}$
= $\longrightarrow \infty $, as it should be because at large
$r$ the result of standard GR has to be reproduced and
$\sigma^2$ has to acquire the value 1.
Taking  $\alpha =(81/8)$ and $\alpha = 8$, the factor 
$\sigma^{2}(\rho )$ 
shows a divergence near $\rho =3/2$ and near $\rho _{+}=%
\frac{(1+\sqrt{5})}{2},$ $\rho _{-}=1$ 
respectively (see Fig. \ref{fig1} and \ref{fig2}). 
In Fig. \ref{fig1} two
lines (dashed and solid) are depicted. The values $\lambda$, 
$\alpha$ and $E$ are kept constant, while the dashed line
is for $\epsilon = 0.1$ (black hole mass ten times $l$) 
and the solid line for  $\epsilon = 0.001$ (a thousand
times $l$). For $\sigma^2$ the behavior is 
regular and finite for $\alpha = 12$, above the critical 
value, even when the mass is only ten times $l$, indicating the (see Fig \ref{fig2}) vanishing of event horizons.

A vanishing $d^2 w$ also indicates that the effective
distance at this point is zero. 

\subsection{Event Horizons}

In order to give an interpretation for the 
divergences of the factor 
$\sigma^{2}(r)$, where it passes zero and tending to 
$-\infty$ (note, that a negative value of $\sigma^2$ is
not allowed, i.e., it corresponds to a non-physical region), 
it is useful to 
determine the position of the event horizon(s) (there may be 
several) for different values 
of $B_4=\alpha m^{4}$. For example, in the 
GR-Schwarzschild solution, the surface $r=2m$ 
is an event horizon. It is also 
a surface of infinite red-shift.
As will be see below, these surfaces are different in the 
pc-Schwarzschild solution \cite{universe}. 
We begin by investigating 
the possible positions of the event horizons 
more explicitly.

A null surface satisfies the equation

\begin{equation}
f(x^{0},x^{1},x^{2},x^{3})=0
~~~,
\label{27} 
\end{equation}
%\color{red}
with $x^0=ct$, $x^1=r$, $x^2=\theta$ and $x^3=\phi$.
%\color{black}

In the case of the Schwarzschild solution in GR 
%\color{red}
the function $f$ for the
%$r-2m,$ so the function $f$ is given by
%\color{black}
event horizon is given by 

\begin{equation*}
f=r-2m=x^{1}-2m
\end{equation*}

Let us turn to pc-Schwarzschild solution, 
for which $g_{\mu \nu }$ is given by

\begin{equation}
g_{\mu \nu }=\left( 
\begin{array}{cccc}
\left( 1-\frac{2m}{r}+\frac{B_4}{6r^{4}}\right) & 0 & 0 & 0 \\ 
0 & -\frac{1}{\left( 1-\frac{2m}{r}+\frac{B_4}{6r^{4}}\right) } & 0 & 0 \\ 
0 & 0 & -r^{2} & 0 \\ 
0 & 0 & 0 & -r^{2}\sin ^{2}\theta%
\end{array}%
\right)
\end{equation}
The $B_4$ can, in principle, acquires any value and 
we will consider the whole range, from 0 to above 
$\frac{81}{8}m^4$.

According to the possible values of $B_4=\alpha m^{4}$, we can distinguished three cases:

\begin{itemize}

\item 1. When $0 \leq \alpha <\frac{81}{8},$ there are two 
event horizons at $r_{\pm ,}$,
for example for \textbf{$\alpha =8$ we get $r_{+}=1.75211m$ and $r_{-}=1.17221m$}, 
%\color{red}
which agrees with \cite{universe}.
%\color{black}

\item 2. For $\alpha =\frac{81}{8}$ the two event horizons $r_{+}$ and $r_{-}$
merge into a single surface at $r_{e}=\frac{3m}{2}$
%\color{red}
as in \cite{universe}. 
%\color{black}
This case
is mainly taken for the applications of pcGR \cite{PPNP}.

\item 3. For $\alpha >\frac{81}{8}$ there are no event 
horizons. 
%and particles
%approaching the center $r=0$ can always return. 

\end{itemize}
\vskip 0.5cm

For all cases, we investigate the
effects within the pc-Schwarzschild geometry. 
The null hypersurface is determined by 
$\left( 1-\frac{2m}{r}+\frac{B_4}{6r^{4}}\right) =0.$
It is clear that, the factor $\sigma ^{2}(r)$ 
diverges at certain points
for $\alpha \leq \frac{81}{8}$ on the two existing
event horizons. For example $\alpha =8$, the factor 
$\sigma
^{2}(r)$ has two divergences at $r_{+}=1.75211m$ and $r_{-}=1.17221m$, respectively
%\color{red}
(see Fig. \ref{fig2}, dashed line). 
%\color{black}
%In
%addition, for $\lambda =0$ the 
%\color{red}
%factor 
%solutions of the equation
%\color{black}
%$\sigma ^{2}(r)=0$ depends on the
%\color{red}
%complicated function of $\sigma^2(r)$ (\ref{25}),
%interrelated with (\ref{25a}), and on the number of event 
%horizons. The $\sigma^2(r)$ tends always for 
%$r \rightarrow 0$ to $-\infty$. 
%number of existing horizons, a number which in turn depends 
%the value of $B$.
%\color{red} 
Below and above each sigunlarity, the $\sigma^2(r)$ function
has to go through zero. According to the position of the
divergences $r_{\pm}$ we call these positions, where 
$\sigma^2(r)$ has a zero, $r_{\pm 1}$ and $r_{\pm 2}$,
respectively.
%\color{black}

In fact, in the
case $B_4<\frac{81}{8}m^{4}$
%\color{red}
(more specifically, the example $\alpha = 8$ is shown)
%\color{black}
the factor $\sigma ^{2}(r)=0$ has five possible values
for $r$ 
%\color{red}
(as explained above), 
%\color{black}
one is near center ($r=0)$ $r_{0}$, two $r_{-1},$ $r_{-2}$ are
%\color{red}
below and above 
%\color{black}
the $r_{-}$, thus $r_{-1}<r_{-}<r_{-2}$ \ and\ two $r_{+1},$ $r_{+2}$ are
%\color{red}
%around
below and above 
%\color{black}
the $r_{+}$, i.e. $r_{+1}<r_{+}<r_{+2}$. 
For $B_4$ = $\frac{81}{8}m^{4}$ 
%\color{red}
($\alpha = \frac{81}{8}$)
%\color{black}
there are three solutions $r$ where 
$\sigma ^{2}(r)=0$. One is always
near the center $r=0$ and two $r_{e1}$ $r_{e2}$ are 
near $r_{e}$,
where the index e refers to the {\it event horizon}. 
For $B_4>\frac{81}{8}m^{4}$
%\color{red}
(more specific $\alpha = 12$),
%\color{black}
there is one value of $\ r$ near the center in which $%
\sigma ^{2}(r)=0.$ Therefore, the factor $\sigma ^{2}(r)$ is always
negative between $(0,$ $r_{0}),$ $(r_{-1},r_{-2})$ and $(r_{+1},$ $r_{+2})$.
%In the case $\lambda \neq 0,$ the factor $\sigma ^{2}(r)=0$ 
%doesn't depend
%on the location of the event horizons. 
As is known, the factor $\sigma ^{2}(r) = 0$
represents a particle with a maximal acceleration 
%\color{red}
$\frac{c^{2}}{l}$,
thus, the zeros of the $\sigma^2(r)$ reveals these positions{of
maximal acceleration.
%\color{black}

\subsection{The effective Potential}

In order to make a direct comparison with the motion in the 
pc-Schwarzschild
geometry possible, we adopt the same procedure
as before. The new action is given by

\begin{equation}
S=\int \sigma ^{2}(r)\left\{ \left( 1-\frac{2m}{r}+\frac{B}{6r^{4}}\right) 
\overset{\cdot }{t}^{2}-\frac{1}{\left( 1-\frac{2m}{r}+\frac{B}{6r^{4}}%
\right) }\overset{\cdot }{r}^{2}-r^{2}(\overset{\cdot }{\theta }^{2}+\sin
^{2}\theta \overset{\cdot }{\phi }^{2})\right\} dw\label{28} 
\end{equation}

The equation of motions become ($\theta =\frac{\pi }{2})$%
\begin{align}
\sigma ^{2}(r)\left( 1-\frac{2m}{r}+\frac{B}{6r^{4}}\right) \overset{\cdot 
}{t}& =E  \label{29} \\
\sigma ^{2}(r)r^{2}\overset{\cdot }{\phi }& =L  \label{30} 
\end{align}

In addition we use $\frac{dw^2}{dw^2}=1$, i.e., 

\begin{equation}
1=\sigma ^{2}(r)\left\{ \left( 1-\frac{2m}{r}+\frac{B}{6r^{4}}\right)
\left( \frac{dt}{dw}\right) ^{2}-\frac{1}{\left( 1-\frac{2m}{r}+\frac{B}{%
6r^{4}}\right) }\left( \frac{dr}{dw}\right) ^{2}-r^{2}\left( \frac{d\phi }{dw%
}\right) ^{2}\right\}  \label{31} 
\end{equation}

Substituting the equations (29) and (30) into 
equation (31), we get

\begin{equation}
\frac{1}{\left( 1-\frac{2m}{r}+\frac{B}{6r^{4}}\right) }\left( \frac{dr}{dw}%
\right) ^{2}=\frac{E^{2}}{\sigma ^{4}(r)\left( 1-\frac{2m}{r}+\frac{B}{%
6r^{4}}\right) }-\frac{L^{2}}{\sigma ^{4}(r)r^{2}}-\frac{1}{\sigma
^{2}(r)}
~~~.
\end{equation}

Solving for $\left(\frac{dr}{dw}\right)^2$, we obtain

\begin{align}
\left( \frac{dr}{dw}\right) ^{2}& =\left( 1-\frac{2m}{r}+\frac{B}{6r^{4}}%
\right) \left[ \frac{E^{2}}{\sigma ^{4}(r)\left( 1-\frac{2m}{r}+\frac{B}{%
6r^{4}}\right) }-\frac{L^{2}}{\sigma ^{4}(r)r^{2}}-\frac{1}{\sigma
^{2}(r)}\right]  \notag \\
& =E^{2}-V_{eff}^{2}(_{r}) 
~~~.
\label{32} 
\end{align}

Comparing both sides we obtain for the effective potential
 
\begin{equation}
V_{eff}^{2}(_{r})=E^{2}-\frac{E^{2}}{\sigma ^{4}(r)}+\left( 1-\frac{2m}{r}+%
\frac{B}{6r^{4}}\right) \left( \frac{L^{2}}{\sigma ^{4}(r)r^{2}}+\frac{1}{%
\sigma ^{2}(r)}\right)  
~~~.
\label{33} 
\end{equation}

In a more traditional approach, one writes
\begin{equation}
\frac{1}{2}\left( \frac{dr}{dw}\right) ^{2}+P_{eff}(r)=\omega
~~~,
\end{equation}
where

\begin{equation}
P_{eff}(r)=\frac{1}{2}(V_{eff}^{2}(_{r})-1)\text{ and }\omega=\frac{1}{2}%
(E^{2}-1)
\end{equation}
In order to plot the 
effective potential, it is preferable to use the 
a-dimensional variables (\ref{dimensionless}) and 
the effective potential  
%\color{red}
becomes, with $\rho = \frac{r}{m}$,
%\color{black}

\begin{equation}
V_{eff}^{2}(_{r})=E^{2}-\frac{E^{2}}{\sigma ^{4}(\rho )}+\left( 1-\frac{2}{%
\rho }+\frac{\alpha }{6\rho ^{4}}\right) \left( \frac{\lambda ^{2}}{\sigma
^{4}(l,\rho )\rho ^{2}}+\frac{1}{\sigma ^{2}(\rho )}\right)  \label{34}
\end{equation}
%\color{red}
Inspecting (\ref{34}), one notes that the potential exhibits
singularities as soon as $\sigma^2(r)$ approaches zero. On
the other hand, when $\sigma^2(r)\rightarrow -\infty$ the 
two last term tends to zero and the effective potential
approaches $E^2$.
%\color{black} 

For 
%\color{red}
a selected set of
%\color{black}
parameter values,
the effective potential is plotted in the Figs. \ref{fig4} to 
\ref{fig15}. In all examples one notes the
important feature that for 
%\color{red}
all $\alpha$ 
%at $\frac{81}{8}$ = 10.125
%\color{black} 
and small $\epsilon$ (large masses) 
the effective potential behaves smoothly, 
showing a repulsive behavior toward small distances. It 
shows a minimum 
%\color{red}
between 1 and 1.5
%\color{black}
which finally increases
toward larger $r$, i.e., it has the typical form of an
attractive "molecular" potential. 

\begin{figure}[H]
\centering
\includegraphics[width=0.60\textwidth]{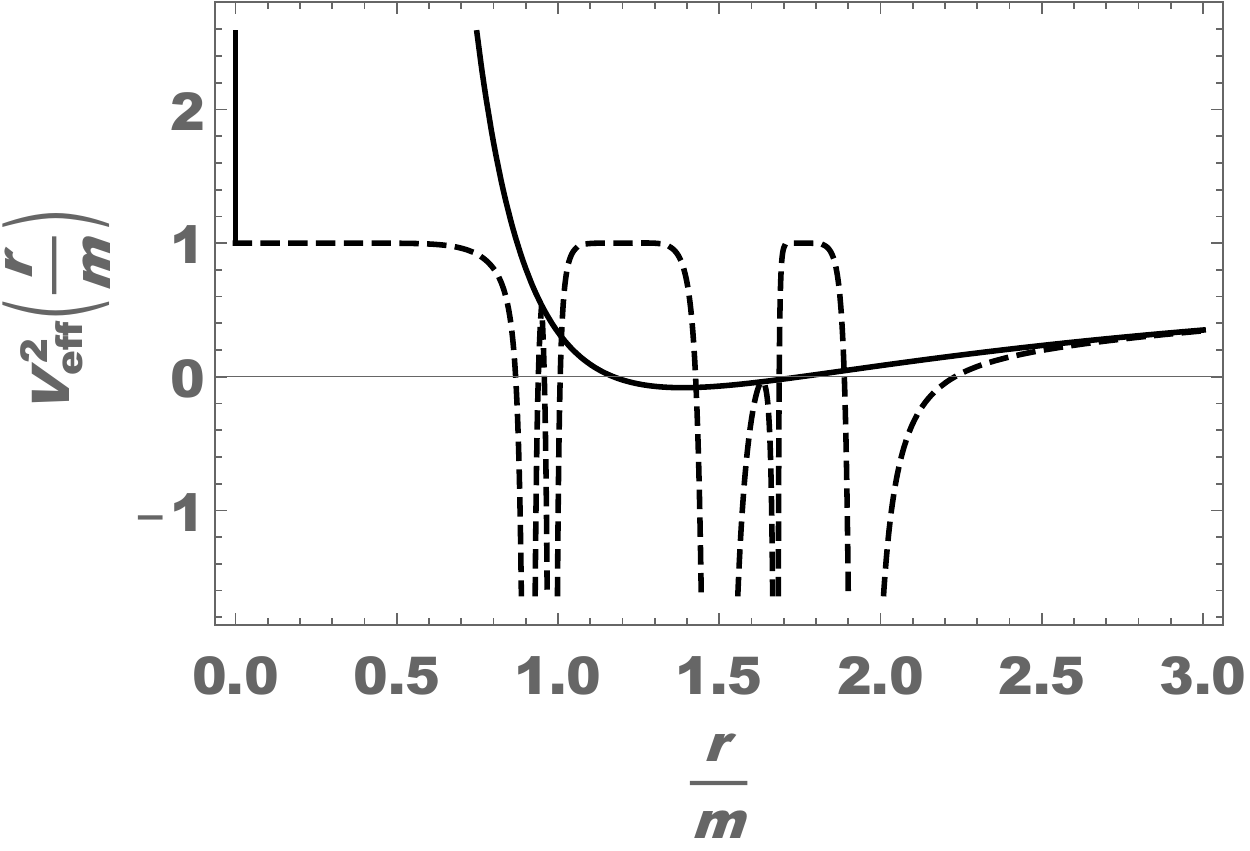} 
\caption{Solid line: the effective potential 
$V_{eff}^{2}(\rho)$ for $\lambda=0$, $\alpha=8,\epsilon=10^{-35}$ and
$E=1$. Dashed line: the effective potential  $V_{eff}^{2}(\rho)$ for $\lambda=0,\alpha=8,\epsilon=0.1$ and $E=1$.\label{fig4}}
\end{figure}

\begin{figure}[H]
\centering
\includegraphics[width=0.60\textwidth]{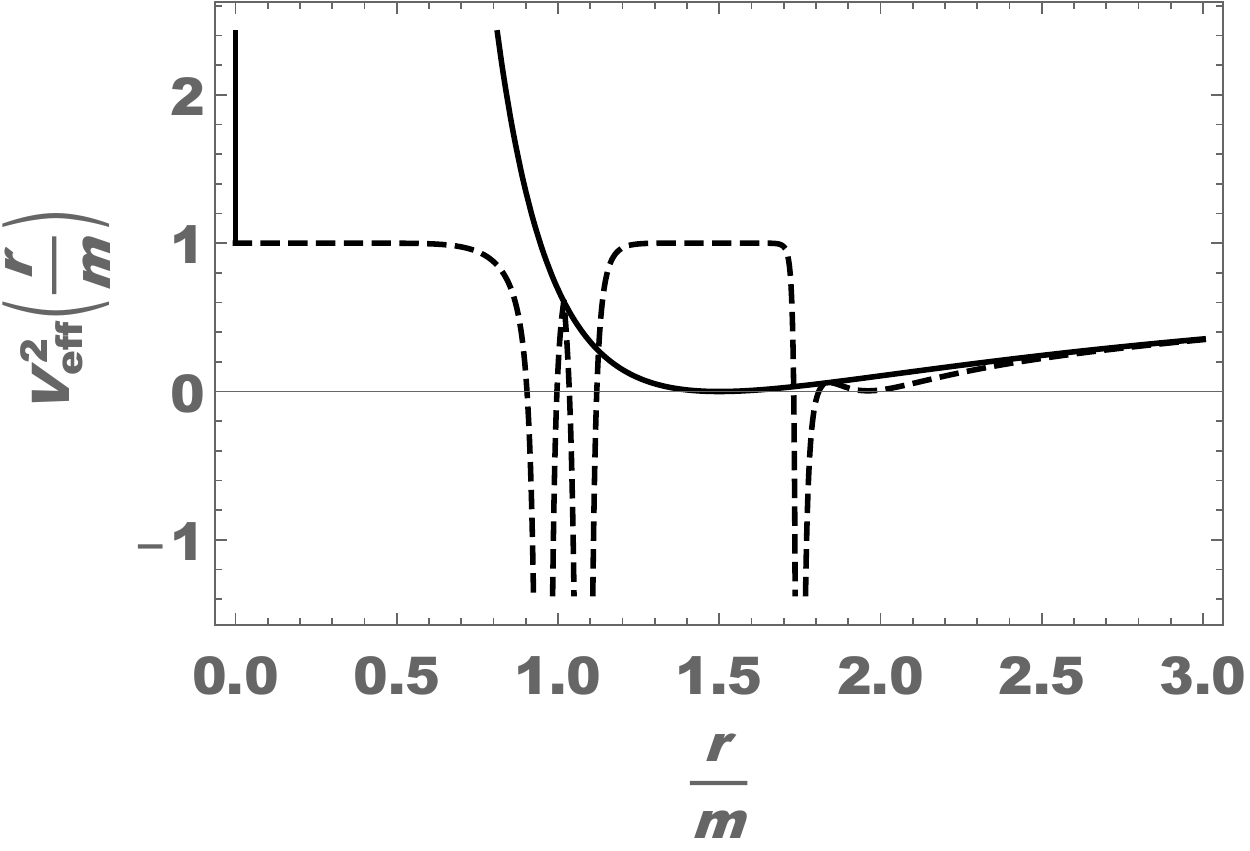} 
\caption{Solid line: the effective potential 
$V_{eff}^{2}(\rho)$ for $\lambda=0$, $\alpha=81/8,\epsilon=10^{-35}$ and
$E=1$. Dashed line: the effective potential  $V_{eff}^{2}(\rho)$ for $\lambda=0,\alpha=81/8,\epsilon=0.1$ and $E=1$.\label{fig5}}
\end{figure}

\begin{figure}[H]
\centering
\includegraphics[width=0.60\textwidth]{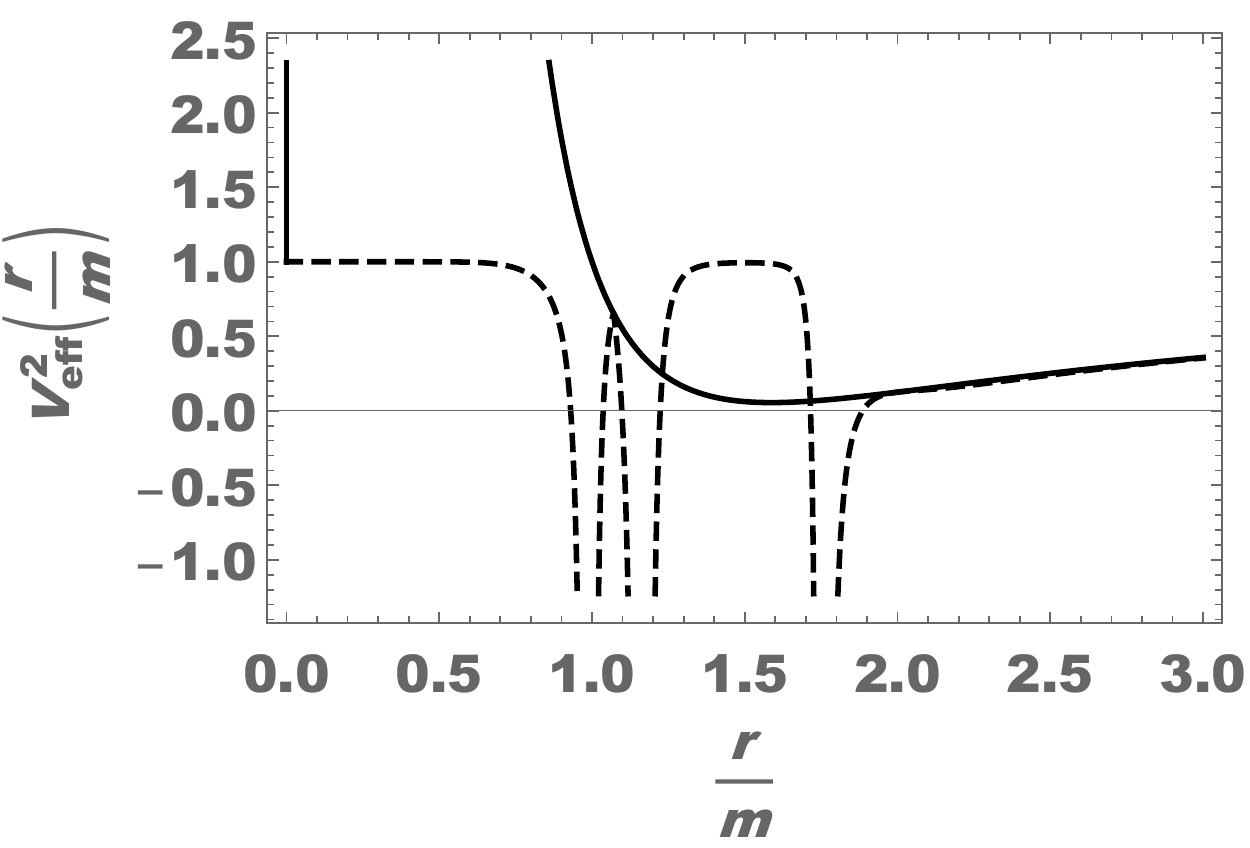} 
\caption{Solid line: the effective potential 
$V_{eff}^{2}(\rho)$ for $\lambda=0$, $\alpha=12,\epsilon=10^{-35}$ and
$E=1$. Dashed line: the effective potential  $V_{eff}^{2}(\rho)$ for $\lambda=0,\alpha=12,\epsilon=0.1$ and $E=1$.\label{fig6}}
\end{figure}

\begin{figure}[H]
\centering
\includegraphics[width=0.60\textwidth]{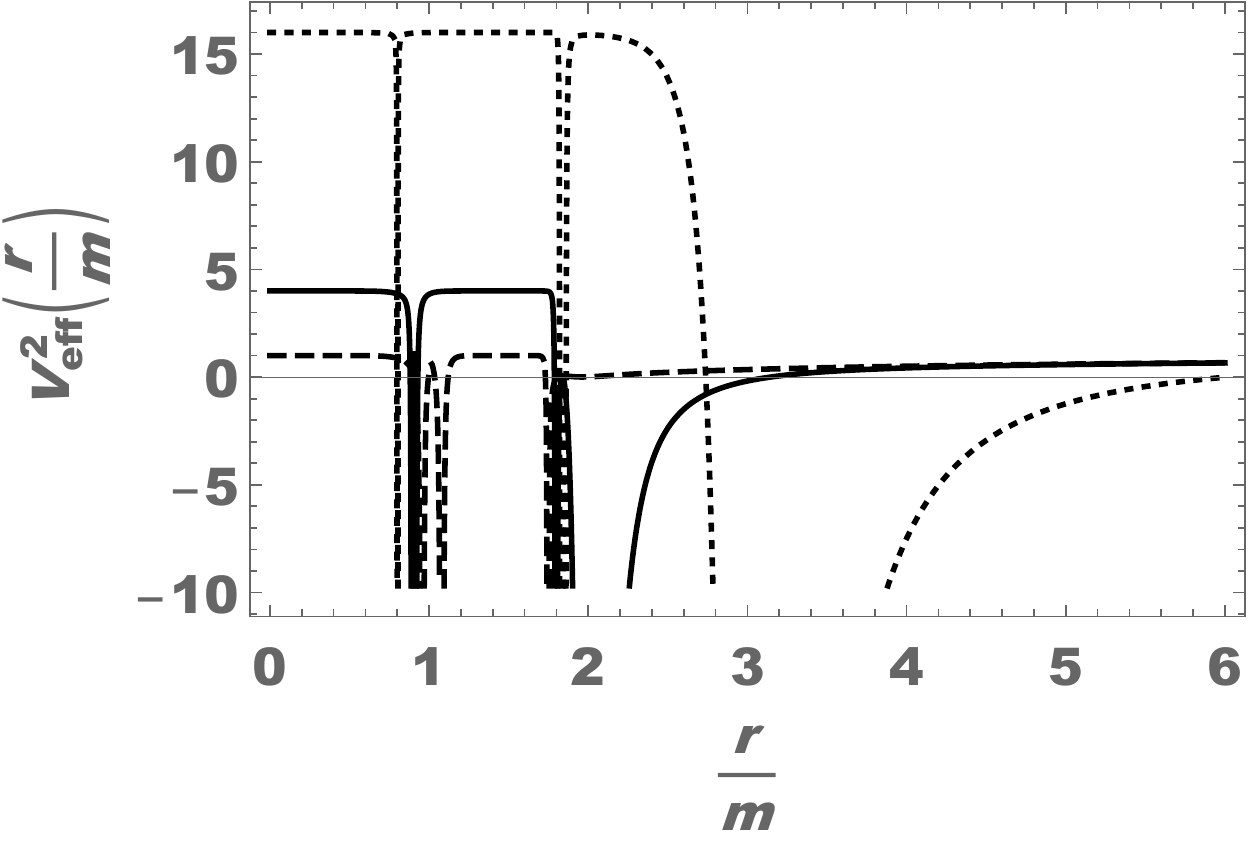} 
\caption{Dashed line: the effective potential 
$V_{eff}^{2}(\rho)$ for $\lambda=0$, $\alpha=81/8,\epsilon=0.1$,
$E=1$. Solid line: the effective potential  $V_{eff}^{2}(\rho)$ for $\lambda=0,\alpha=81/8,\epsilon=0.1$ and $E=2$ and Dotted line: the effective potential  $V_{eff}^{2}(\rho)$ for $\lambda=0,\alpha=81/8,\epsilon=0.1$ and $E=4$.\label{fig6.1}}
\end{figure}

\begin{figure}[H]
\centering
\includegraphics[width=0.60\textwidth]{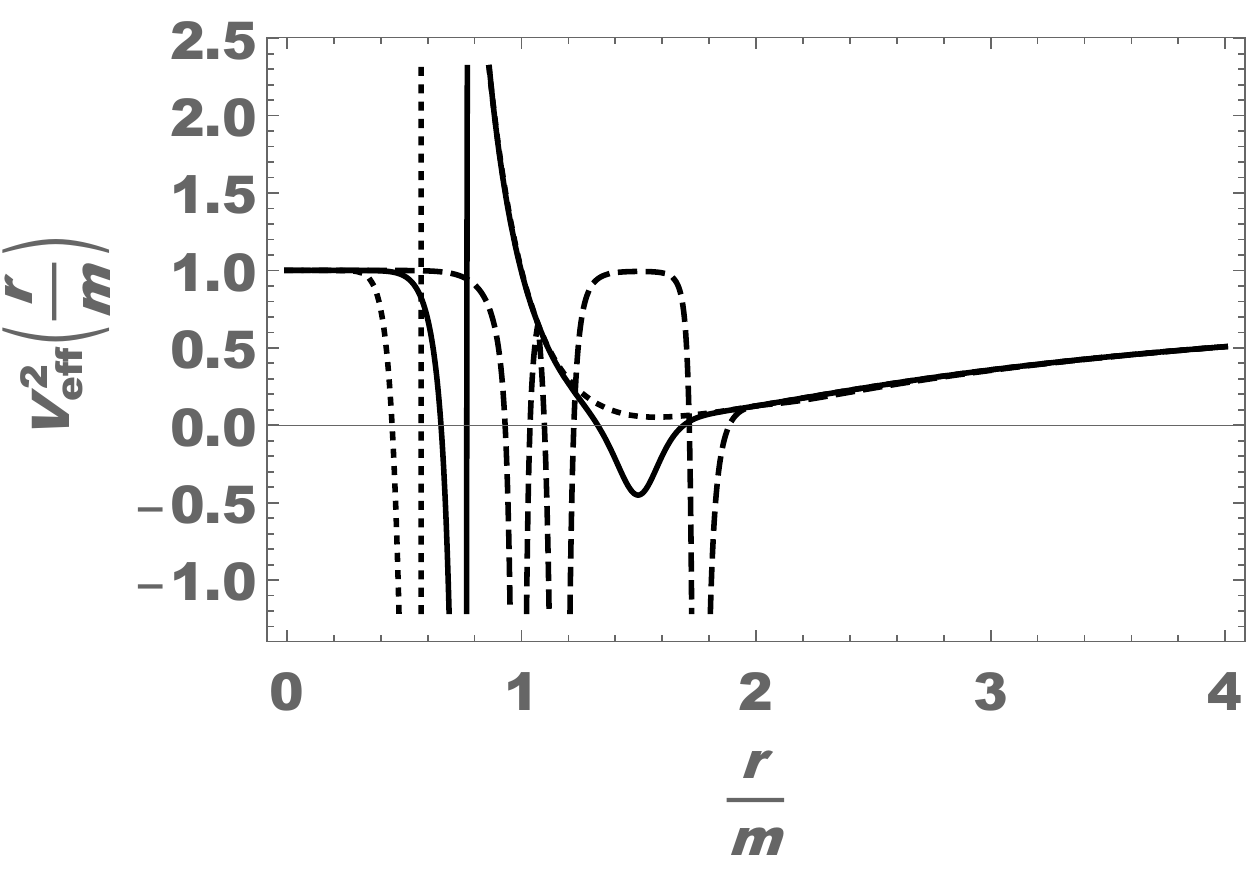} 
\caption{Dashed line: the effective potential 
$V_{eff}^{2}(\rho)$ for $\lambda=0$, $\alpha=12,\epsilon=0.1$ and
$E=1$. Solid line: the effective potential  $V_{eff}^{2}(\rho)$ for $\lambda=0,\alpha=12,\epsilon=0.01$ and $E=1$. Dotted line: the effective potential  $V_{eff}^{2}(\rho)$ for $\lambda=0,\alpha=12,\epsilon=0.001$ and $E=1$. \label{fig7}}
\end{figure}

\begin{figure}[H]
\centering
\includegraphics[width=0.60\textwidth]{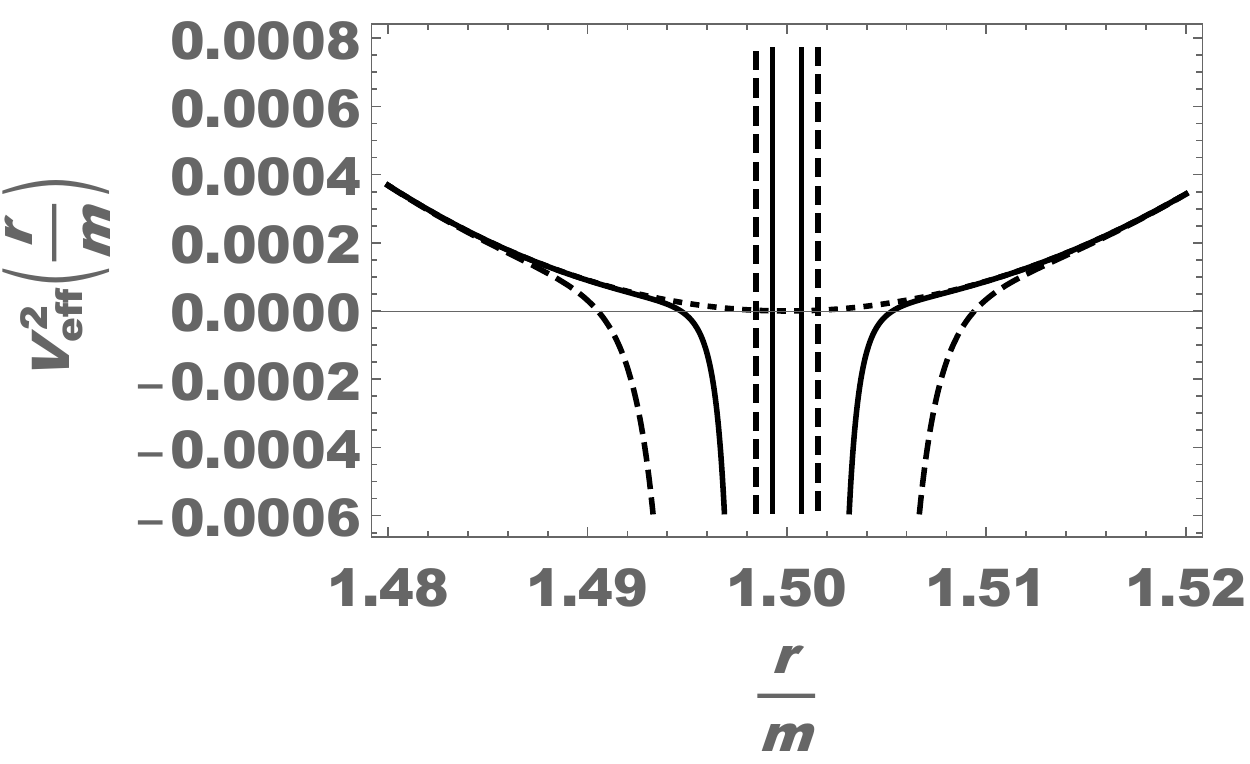} 
\caption{
%\color{red}
A zoom of the effective potential near $\frac{r}{m}=1.5$, with
the intention to note the existence of a potential barrier
still for rather large black hole masses. For very large 
masses, e.g. $\epsilon = 10^{-20}$. the barrier 
%\color{red}
seems to have vanished. But effects are present 
up to $\epsilon=10^{-18}$ (not shown here).
%\color{black}
%\color{black}
Dashed line: the effective potential 
$V_{eff}^{2}(\rho)$ for $\lambda=0$, $\alpha=81/8,\epsilon=10^{-8}$ and
$E=1$. Solid line: the effective potential  $V_{eff}^{2}(\rho)$ for $\lambda=0,\alpha=81/8,\epsilon=10^{-9}$ and $E=1$. Dotted line: the effective potential  $V_{eff}^{2}(\rho)$ for $\lambda=0,\alpha=81/8,\epsilon=10^{-20}$ and $E=1$. \label{fig7a}}
\end{figure}

%\begin{figure}[H]
%\centering
%\includegraphics[width=0.60\textwidth]{effecpoten6.pdf} 
%\caption{Dashed line: the effective potential 
%$V_{eff}^{2}(\rho)$ for $\lambda=90$, $\alpha=12,\epsilon=0.1$ and
%$E=10$. Solid line: the effective potential  $V_{eff}^{2}(\rho)$ for $\lambda=90,\alpha=8,\epsilon=0.1$ and $E=10$. Dotted line: the effective potential  $V_{eff}^{2}(\rho)$ for $\lambda=90,\alpha=81/8,\epsilon=0.1$ and $E=10$. \label{fig9}}
%\end{figure} 

\begin{figure}[H]
\centering
\includegraphics[width=0.60\textwidth]{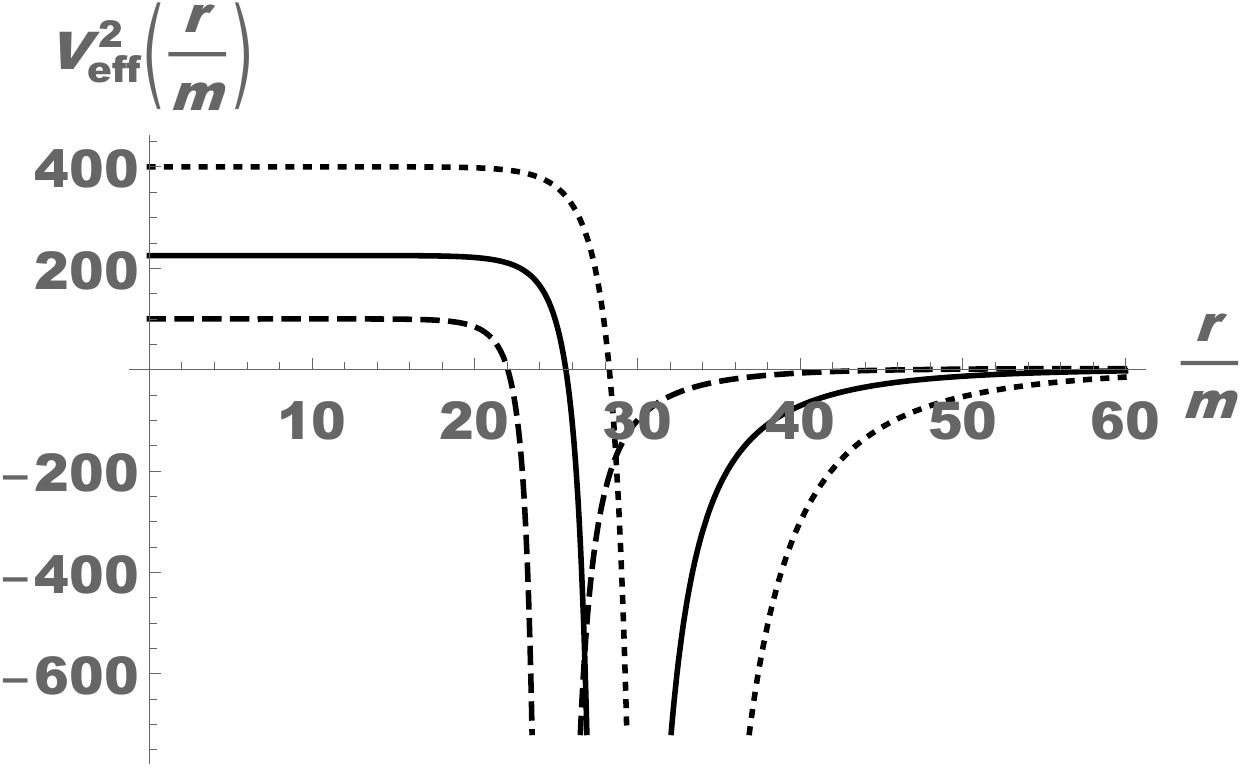} 
\caption{Dashed line: the effective potential 
$V_{eff}^{2}(\rho)$ for $\lambda=90$, $\alpha=12,\epsilon=0.1$ and
$E=10$. Solid line: the effective potential  $V_{eff}^{2}(\rho)$ for $\lambda=90,\alpha=12,\epsilon=0.1$ and $E=15$. Dotted line: the effective potential  $V_{eff}^{2}(\rho)$ for $\lambda=90,\alpha=12,\epsilon=0.1$ and $E=20$. 
\label{fig10}}
\end{figure} 

%\begin{figure}[H]
%\centering
%\includegraphics[width=0.60\textwidth]{effecpoten8.pdf} 
%\caption{Dashed line: the effective potential 
%$V_{eff}^{2}(\rho)$ for $\lambda=90$, $\alpha=12,\epsilon=0.01$ %and
%$E=10$. Solid line: the effective potential  $V_{eff}^{2}(\rho)$ for $\lambda=90,\alpha=12,\epsilon=0.01$ and $E=15$. Dotted line: the effective potential  $V_{eff}^{2}(\rho)$ for $\lambda=90,\alpha=12,\epsilon=0.01$ and $E=20$. \label{fig11}}
%\end{figure} 

%\begin{figure}[H]
%\centering
%\includegraphics[width=0.60\textwidth]{effecpoten9.pdf} 
%\caption{Dashed line: the effective potential 
%$V_{eff}^{2}(\rho)$ for $\lambda=90$, $\alpha=12,\epsilon=0.001$% and
%$E=10$. Solid line: the effective potential  $V_{eff}^{2}(\rho)$ for $\lambda=90,\alpha=12,\epsilon=0.001$ and $E=15$. Dotted line: the effective potential  $V_{eff}^{2}(\rho)$ for $\lambda=90,\alpha=12,\epsilon=0.001$ and $E=20$. \label{fig12}}
%\end{figure} 

%\begin{figure}[H]
%\centering
%\includegraphics[width=0.60\textwidth]{effecpoten11.pdf} 
%\caption{Dashed line: the effective potential 
%$V_{eff}^{2}(\rho)$ for $\lambda=90$, $\alpha=(81/8,\epsilon=0.0001$ and
%$E=10$. Solid line: the effective potential  $V_{eff}^{2}(\rho)$ for $\lambda=90,\alpha=81/8,\epsilon=0.0001$ and $E=15$. Dotted line: the effective potential  $V_{eff}^{2}(\rho)$ for $\lambda=90,\alpha=81/8,\epsilon=0.0001$ and $E=20$. \label{fig13}}
%\end{figure} 

\begin{figure}[H]
\centering
\includegraphics[width=0.60\textwidth]{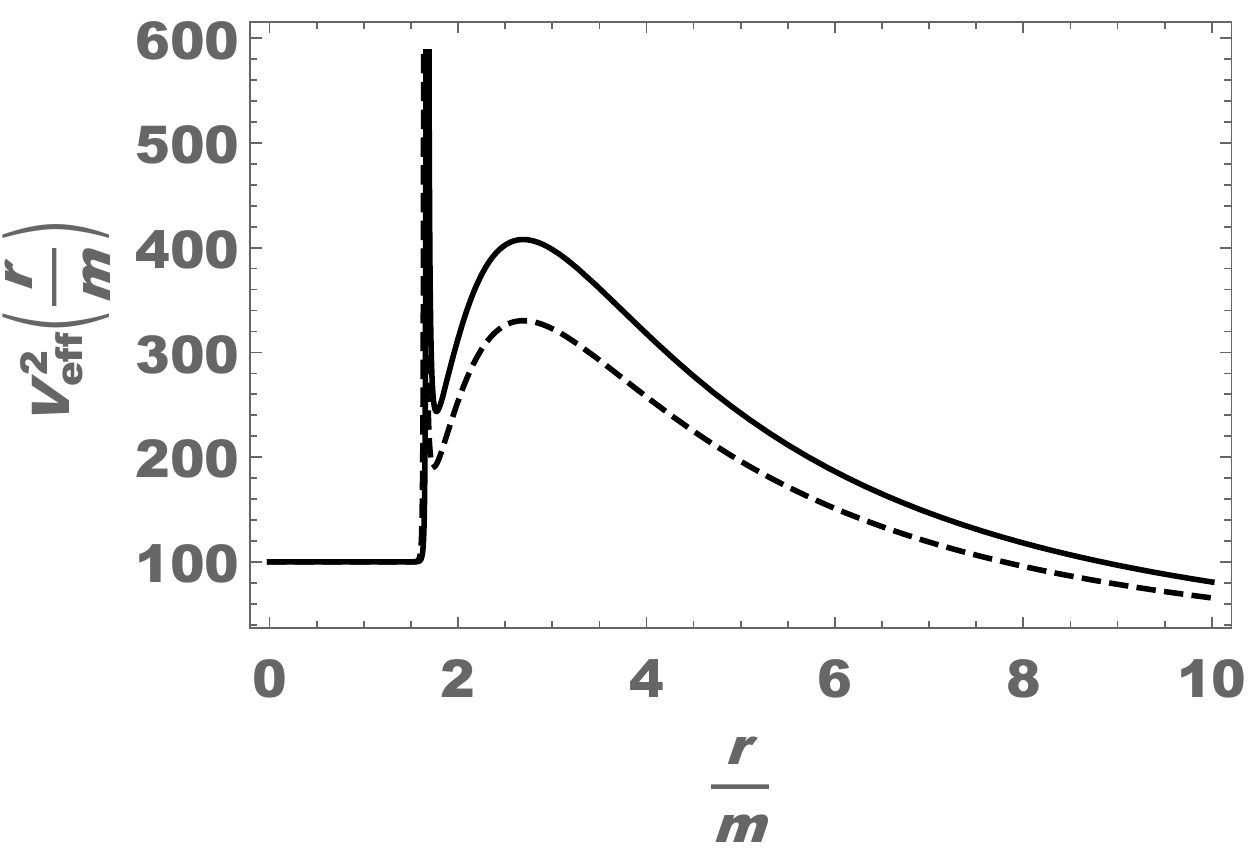} 
\caption{Dashed line: the effective potential 
$V_{eff}^{2}(\rho)$ for $\lambda=90$, $\alpha=(81/8,\epsilon=0.000001$ and $E=10$. Solid line: the effective potential  $V_{eff}^{2}(\rho)$ for $\lambda=100,\alpha=12,\epsilon=0.000001$ and $E=10$. 
%\color{red}
For $\frac{r}{m} \rightarrow 0$ the effective potential approaches $E^2=100$.
%\color{black}
\label{fig14}}
\end{figure} 

\begin{figure}[H]
\centering
\includegraphics[width=0.60\textwidth]{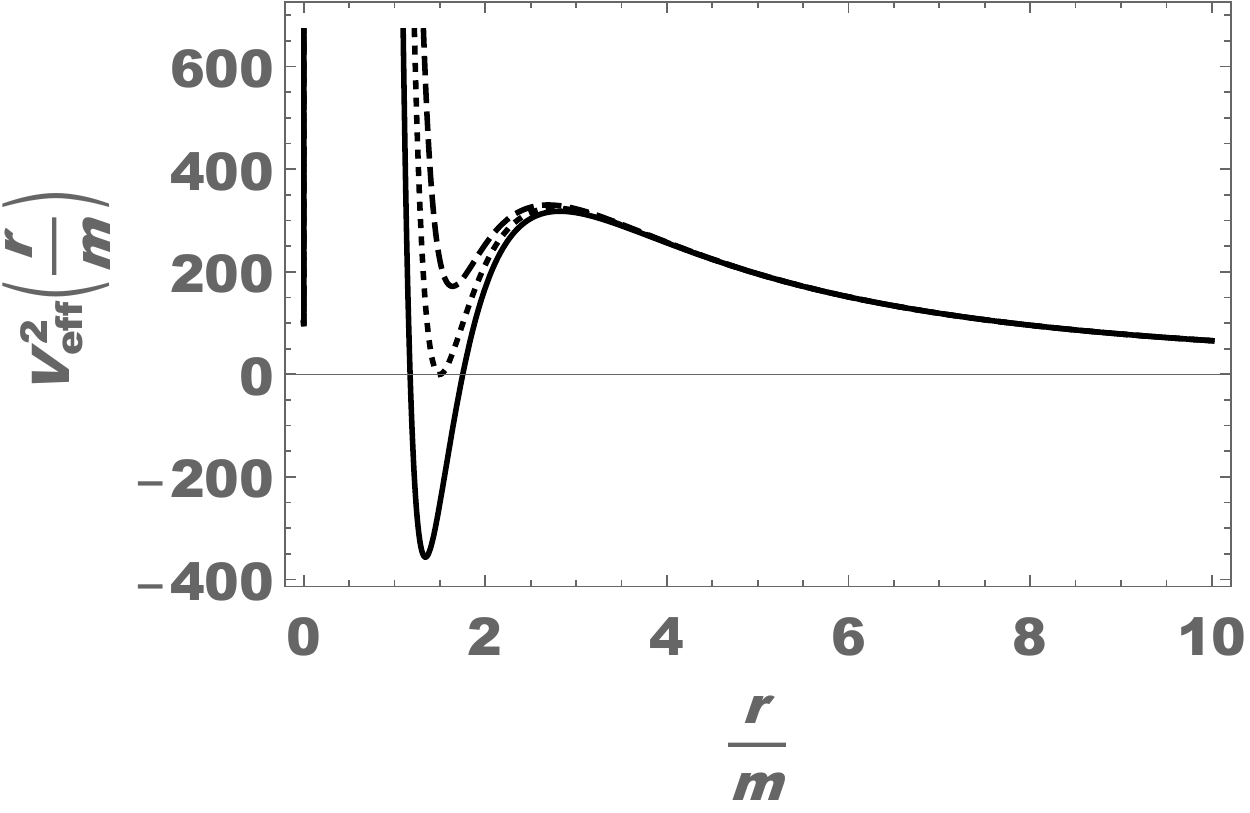} 
\caption{Dashed line: the effective potential 
$V_{eff}^{2}(\rho)$ for $\lambda=90$, $\alpha=12,\epsilon=10^{-35}$ and
$E=10$. Solid line: the effective potential  $V_{eff}^{2}(\rho)$ for $\lambda=90,\alpha=8,\epsilon=10^{-35}$ and $E=10$. Dotted line: the effective potential  $V_{eff}^{2}(\rho)$ for $\lambda=90,\alpha=81/8,\epsilon=10^{-35}$ and $E=10$. 
%\color{red}
The nearly vertical line at small $\frac{r}{m}$ is the potential
coming down from large values and approaching $E^2=100$ at
$\frac{r}{m}=0$. Due to the poor resolution, the bending of
the curve near 0 is not resolved.
%\color{black}
\label{fig15}}
\end{figure} 

The following discussion is easier to understand, noting that
the effective potential in (\ref{34}) has two important terms:
the term $-E^2/\sigma^4$ becomes dominant when the $\sigma^2$
%\color{red}
%exhibits a pole, where it 
tends to zero.
%\color{black}
In this case,
the potential goes to $-\infty$. 
%The term
%$\left( 1-\frac{2}{\rho }+\frac{\alpha }{6\rho ^{4}}\right)$
%= $g_{00}$
%($\rho = \frac{r}{m}$) becomes singular for $r \rightarrow 0$,
%where it tends to $+\infty$.
%\color{red}
When $\sigma^2(r)$ tends to $-\infty$, the only term which
survives is the first one, namely $E^2$, which the effective 
potential finally acquires.
%\color{black}

In Figs. \ref{fig4} and \ref{fig5} we studied the effect of 
changing $\alpha$ from 8, below the critical limit, to 
$\alpha=\frac{81}{8}$, the critical limit,
%\color{red}
and $\alpha = 12$, above the critical limit. 
In these
%both 
figures
%\color{black}
the solid line represents a macroscopic black hole and
it shows a typical behavior of an attractive, molecular 
type of potential. In reference to it, in Fig. \ref{fig4}
for $\alpha$ we used the value of 8, in Fig. \ref{fig5}
the value $\frac{81}{8}$ and in Fig. \ref{fig6} the
$\alpha$ is above the critical value. We see that contrary to 
\cite{feoli1999} several potential barriers appear. 
%\color{red}
This is due to the behavior of the $\sigma^2(r)$-function
which exhibits several zeros and also
%\color{black}
the potential
exhibits various 
%\color{red}
zeros
%singularities 
where $\sigma^2(r)$ 
%\color{black}
tends to $-\infty$.
%The solid nearly vertical line near $\frac{r}{m}=0$ is
%artificial and is the result of the metric term
%$g_{00}$ which tends to $+\infty$ for $r \rightarrow 0$. 
We also see that when the 
$\alpha$ changes, also the number of singularities change.

Interesting is how the height of the barrier changes with
increasing energy $E$ of the particle, see
Fig. \ref{fig6.1}. As in \cite{feoli1999} the
height of the barrier increases with $E$, indicating that
the particle is always inhibited to enter the area below the first position of the barrier.
%\color{red}
Note, that near the singularity of $\sigma^2(r)$ the
effective potential barrier approaches $E^2$ (see the
discussion above), thus, the barrier increases with $E^2$.
%\color{black}

In Fig. \ref{fig7} we studied the effect of lowering 
$\epsilon$ (increasing the mass of the black hole), from
$\epsilon =0.1$ on, 
which is ten times the minimal value possible.
We see, that with increasing mass (smaller $\epsilon$)
the barrier structure with its singularities vanish. the effective potential for large black hole masses
%($10^8$, $10^9$ and $10^{20}$ times the minimal length)
is plotted in the vicinity of $\rho = \frac{3}{2}$,
where the event horizon still exists. We note a sharp barrier,
which is the result of introducing a minimal length. Thus,
no particle can pass this point, while without the minimal
length a particle can fall in. Here we see an important effect
in taking into account the presence of a minimal length.

%\color{red}
In Fig. \ref{fig7a} a zoom to near $\frac{r}{m}=1.5$
of the effective potential is
depicted, for the critical $\alpha$ value of $\frac{81}{8}$.
Without a minimal length, the event horizon is at 
$\frac{r}{m}=\frac{3}{2}$, where according to our calculations still a repulsive barrier
is noted up to at least $\epsilon = 10^{-18}$
($10^{18}$ times the Planck length), which is already of macroscopic size, compared to the scale of an elementary
particle! This 
%\color{red}
%barrier is not present anymore for $\epsilon = 10^{-20}$.
barrier is not seen anymore
for $\epsilon = 10^{-20}$. That the effective potential approaches zero
at $r=\frac{3}{2}m$ for $\alpha = \frac{81}{8}$ and has
there a minimum. For $\alpha = 8$, at
the position of the event horizons
$r_+=1.75211m$ and $r_-=1.17221m$, the potential is zero but does 
not acquire a minimum there. These observations 
can be understood from the expression of the effective potential in 
Eq. \ref{33}: The factor $\left( 1-\frac{2m}{r}+
\frac{B}{6r^{4}}\right)$ is zero at the $r$ positions mentioned
and the effective potential reduces to 
$E^2 - \frac{E^2}{\sigma^2}$. The $\sigma^2 (r)$ acquires there 
the value of 1, thus, the two terms cancel and the effective
potential is zero. This is a unique behavior at the
values of $\alpha$  mentioned, which disappears when
the $\alpha$-value is larger than its limiting value, 
because then the factor in from of 
the last term in (\ref{33}) is not zero anymore. 
%\color{black}
%\color{black}

In Fig. \ref{fig10} the angular momentum is now 
different from zero. Increasing the Energy of the particle,
with $\alpha = 12$, the main effect is an increase of 
the barrier, i.e., similar as discussed before. Also,
introducing a large angular momentum smooths out the singular structure of the potential.

Finally, in Figs. \ref{fig14} and \ref{fig15} nearly macroscopic 
black holes (very small $\epsilon$) 
are studied, with a large angular momentum.
Increasing $\alpha$, only lifts the potential to
higher values, which is due to the dark energy alone, no
$l$-dependence is observed. 
%\color{red}
According to (\ref{34}) the effective potential always tends to
$E^2$ at $r=0$. This is because the $\sigma^2(r)$-function
goes to $-\infty$, which eliminates the last terms in
(\ref{34}). Approaching $r=0$, the potential comes from very
large values, which can be seen as a nearly vertical
line in the figures.
%Note that the potential goes to
%infinity near $\frac{r}{m} = 0$, which is an 
%artifact.
%\color{black}
 
%\begin{figure}[!h]
%\centering
%\includegraphics[width=0.45\textwidth]{effecpoten6.pdf} 
%\caption{red line: the effective potential 
%$V_{eff}^{2}(\rho,\epsilon)$ for $\lambda=90$, $\alpha=12,%\epsilon=10^{-35}$ and
%$E=10$. Blue line: the effective potential  $V_{eff}^{2}(\rho,\epsilon)$ for $\lambda=90,\alpha=8,\epsilon=10^{-35}$ and $E=10$. Green line: the effective potential  $V_{eff}^{2}(\rho,\epsilon)$ for $\lambda=90,\alpha=81/8,\epsilon=10^{-35}$ and $E=10$. \label{fig7}}
%\end{figure}
 
%\begin{figure}[H]
%\centering
%\includegraphics[width=0.45\textwidth]{effecpoten5.pdf} 
%\caption{red line: the effective potential 
%$V_{eff}^{2}(\rho,\epsilon)$ for $\lambda=0$, $\alpha=12,\epsilon=0.02$ and
%$E=1$. Blue line: the effective potential  $V_{eff}^{2}(\rho,\epsilon)$ for $\lambda=0,\alpha=13,\epsilon=0.02$ and $E=1$. Green line: the effective potential  $V_{eff}^{2}(\rho,\epsilon)$ for $\lambda=0,\alpha=14,\epsilon=0.002$ and $E=1$. 
%\label{fig8}}
%\end{figure} 

\subsection{Minimal mass of an accelerating object $A_{m}=c^{2}/l$}

In \cite{hawking} its is argued that there exists a
value for the minimal mass of a black hole, equal to the
Plank mass. The main argument is 
that causality has to be maintained and 
that for smaller mass 
quantum effects set in, thus the classical theory is not valid
any more  \cite{21,feoli1999}. 
In this sub-section we show that 
due to the minimal length the limit 
of a minimal black hole mass 
increases.

The new mass scale is obtained when $m$ is
equal to the minimal length, which we will put here as
the {\it Planck length} ($m=l$), or
%\color{red}
$m=l=10^{-35}$meter.
%\color{black} 
%%%ELIMINATED DUE TO REPETITION start
%Lower masses are not allowed
%because it would else correspond to a smaller length as $l$
%and would violate causality \cite{feoli1999,brandt1989}. 
%%%end

%For the sake of transparency, in this sub-section 
%we will relate $m$ to the standard units, denoting by $m$
%the mass in units of length and by $M$ the mass in $kg$: 

The Planck mass is determined as

\begin{eqnarray*}
M_{P} &=&\sqrt{\frac{\hbar c}{k}}=\left( \frac{0.105457\times
10^{-33}\times 3\times 10^{8}}{6.67384\times 10^{-11}}\right) ^{\frac{1}{2}}
\\
&=&0.217726\times 10^{-7}\text{ kg}
\end{eqnarray*}%

Setting  $m=l$ implies $\epsilon = 1$, which in turn leads to
%\color{red}
(using $m=\frac{k M_{\rm object}}{c^2}$, where
$M_{\rm object}$ is the mass in kg)
%\color{black}

\begin{eqnarray}
\epsilon &=&\frac{l}{m}=\frac{lc^{2}}{kM_{\text{object}}}
\label{mass1}
\end{eqnarray}

%\color{red}
Setting now the limit $\epsilon = 1$ and solving for 
$M_{\rm object}$, we obtain for the mass

\begin{eqnarray}
M_{\text{object}} &=&\frac{lc^{2}}{k}=\frac{9\times 10^{16}\times
6.62606\times 10^{-34}}{6.67384\times 10^{-11}}=8.93557\times 10^{-7}\text{
kg}
~~~.
\label{mass2}
\end{eqnarray}
5\color{black}

The ratio of $M_{\rm object}$ to $M_{P}$ is

\begin{equation}
\frac{M_{{\rm object}}}{M_{P}}=41
~~~.
\label{26} 
\end{equation}

As a consequence, the minimal mass of an object, 
corresponding to the minimal length scale,
is $41$ times larger than the Planck mass, a more stringent
limit than the one proposed by S. Hawking \cite{hawking}.

\section{Discussion and Conclusions}
\label{section4}

In this contribution we investigated on the effects of a minimal
length scale parameter in an extended version of 
{\it General Relativity}, called {\it pseudo-complex
General Relativity}, which is equivalent to introducing a
$r$-dependent mass \cite{gw}.
%\color{red} 
Thus, the pc-GR can be seen as a representative other models 
which try to extend General Relativity.
%\color{black}
In past applications of pcGR the
minimal length was ignored, but here we showed that it
generates important differences for small black holes, 
which we will resume in what follows. It is important to
stress that {\it the minimal length is treated as a parameter
and as a consequence the Lorentz symmetry is maintained}.
This represents a great advantage to theories treating
the minimal length as a physical length, violating the
Lorentz symmetry resulting in laborious, involved theories.
%\color{red}
One may ask if the use of a minimal length {\it as a parameter}
is physical or not. In spite of this question, even when the
theory presented is thought as an effective theory with a 
minimal length, the consequences discussed in this manuscript
can serve as an orientation for more general theories, thus,
be very useful.
%\color{black}

A minimal length is related to a maximal acceleration.
Its effect is cast into a conformal factor 
$\sigma^2(r )$ of the metric.
When the acceleration of a particle's  
tends to its maximal value, then the metric correction factor 
$\sigma ^{2}(r )\longrightarrow 0$. 
At large distances the classical potential of GR
is recovered, i.e., 
$V_{eff}^{2}(r)\longrightarrow 1$ as $r
\longrightarrow \infty$. 

As one main result, we showed that the effects of the
minimal length can 
%\color{red}
%only
mainly 
%\color{black}
be noted when the mass is of the
order of the Planck mass and a few orders larger than it.
%\color{red}
%{\it For macroscopic black holes no effects can be seen.}
{\it For macroscopic black holes also some effect can be seen
when $\alpha = \frac{81}{8}$}, up to $\epsilon = 10^{-18}$,
corresponding to a length of $10^{-15}$fm,
{\it which is of the order of the size of a hadron.}
%\color{black}

But there is also an effect in pcGR due to the coupling
of the central mass to the size of the dark energy in the
vicinity of a black hole.
Depending on the value of $B_4=bm^4$, we can distinguish
three cases, given by particular ranges of the parameter
$B_4$, where we only 
%\color{red}
%consider 
resume the results 
%\color{red}
mainly
%\color{black}
for
%\color{black}
zero angular momentum 
($\lambda = 0$),
%\color{red}
though, also some remarks on $\lambda > 0$ are included:
%\color{black}

\begin{itemize}

\item \textbf{case a:} $B_4<81m^{4}/8$

The range of $\epsilon =(l/m)$ is divided into two regions,
the first is $\epsilon_0 \leq \epsilon \leq 1$, where
$\epsilon_0$ marks the value above which no singularity in
$\sigma^2$ appears. In the second case 
($\epsilon < \epsilon_0$) the conformal factor $\sigma^2$
s-hows singularities, as can be appreciated in Fig. 1 and 2,
where the factor shows three and five points where $\sigma^2$
passes to zero to infinite negative values. 
The region where $\sigma^2$ is negative is unphysical and has 
to be excluded. The number and position of these singularities
vary with $\epsilon$. The exact value of $\epsilon_0$ cannot
be given but only estimated.
 
%\color{red}
We have two cases to consider. For the 
first one $\epsilon_0 \leq \epsilon \leq 1$, 
%\color{black}
the effective potential tends to 
$E^{2}$ near $r_{\pm }$
and the center $r=0$. Moreover, the effective potential diverges when $\sigma
^{2}(\rho )=0$ in the range
$r_{0}<r_{-1}<r_{-2}<r_{+1}<r_{+2}$. 
(The $r_m$ values refer to the position 
where $\sigma^2(r)$ passes through zero, see
the main text.)
%For example, one sees in 
%Fig. \ref{fig2} that we obtain $r_{0}=0.539192,$ $r_{-1}=0.802579,$ $r_{-2}=1.14495,$ $%
%r_{+1}=1.35233,$ and $r_{+2}=1.91432$. 
The divergences appear as a
potential barrier near $r_{\pm }$ and at the center ($r=0$).
The potential barrier also increases with larger
values of $E$ and,
as a consequence, the
incoming massive particle of energy $E$ would never 
falls into the black hole, i.e., the accretion of mass to the 
mini-black hole is stopped.
Fig. \ref{fig4} and subsequent figures 
show that the effective potentials can not be distinguished from
each other at infinity. 

For $\lambda \neq 0,$ the effective potential
still has the value of 
$E^{2}$ near the center $r=0$ and 
to the horizons $r_{+}$ and 
$r_{-}$. There is a 
critical values of $E$, which
determines the sign of effective potential; it becomes 
positive when 
$E^{2}<\frac{\lambda ^{2}}{\rho _{c}^{2}}(1-\frac{2}{\rho _{c}}+\frac{\alpha }{6\rho _{c}^{4}})$
%\color{red}
where $\rho_c$ is the value of $\rho$ when coming from
$\rho =0$ the effective potential becomes negative (see
Figs. \ref{fig10}-\ref{fig15}). 
%\color{black}
At a higher value 
of the critical energy, there are two
singularities in which the factor $\sigma ^{2}(r )=0.$

The second case is for 
%\color{red}
$\epsilon \ll 1$, which corresponds to the 
classical limit, i.e., the mass, in length 
units, of the black hole is much bigger than the 
minimal length.
%\color{black} 
This means that $\sigma ^{2}(r)\longrightarrow 1.$ As 
consequence, the potential barrier in this case
disappears. Due to this, the effects of the 
minimal length can be neglected.

\item \textbf{case b:} $B_4=81m^{4}/8$

In this case, according the
value of $\epsilon =(l/m)$, we can also distinguish two 
ranges: In the first one, we have 
$\epsilon _{0}$ $\leq \epsilon \leq 1$ (here $\epsilon _{0}$ represents the
value of $\epsilon $ for which it makes the transition 
from the first to the second
range). The effective potential has three singularities, one 
near zero and
two degeneraste ones at $\rho _{e}$ = $\frac{r_e}{m}$ = 
$\frac{3}{2}$ (the index $e$ refers to the event horizon)
but the effective potential has the
maximum value $E^{2}$ for 
$r _{e}=\frac{3m}{2}$ and near the center $r=0.$
These singularities act as a potential barrier 
(see, for example, Fig. \ref{fig4}). When $\alpha =\frac{81}{8}$
the effective potential for a large mass black hole still
shows a repulsive barrier at $\rho = \frac{3}{2}$, which
inhibits the falling in of particles, contrary to the case
when no minimal length is taken into account, where the 
particle can pass the horizon. 
%\color{red}
It is noticeable that
at the position of the event horizon ($\rho_e = \frac{3}{2}$)
there is still a very large barrier for relative large black
hole masses, at least up to $10^{18} l$ and the penetration
of a particle is suppressed. For much larger black hole
%\color{black}
%\color{red}
%masses this barrier disappears.
of $\epsilon = 10^{-20}$
masses this barrier seems to disappear.
An $\epsilon = 10^{-18}$ corresponds
to a length of $10^{-15}$cm, which implies the possibility that
effects of the minimal length could be seen there, if
$\alpha = \frac{81}{8}$ and these small effect can be
measured.
%\color{black}

For the second range, we consider 
$\epsilon _{0}$ $\geqslant \epsilon ,$ here $\epsilon =(l/m)$ is
very small
(i.e., the black hole is of macroscopic size), 
this means $\sigma ^{2}(r )\longrightarrow 1,$ 
in this case
the effective potential has no singularities. This implies
that there is also no
potential barrier. It is clear that the effect of a minimal 
length is large
when $l\approx m$, i.e., for microscopic black holes.

Fig. \ref{fig1} \ shows that $\sigma ^{2}(r )$ diverges 
for the same values of 
$\rho _{e}$. The $\sigma ^{2}(\rho )$ always appears in the 
denominator of
the effective potential (6) which thus equals $E^{2}$.
This means that the
velocity of any incoming particle becomes zero 
(see equation (5)). The
same behavior is seen in Fig. \ref{fig4}. 
For the radial motion, there are always
divergencies (i.e., singularities) 
produced by the zero of the factor 
$\sigma^{2}(\rho )$.

\item \textbf{case c:} $B_4>81m^{4}/8$

For this case,  
in the range  
$\epsilon _{0}$
$\leq \epsilon \leq 1$, the effective potential has also a barrier in the potential, 
see Fig. \ref{fig4} for $\lambda =0$. In
addition, for $\alpha =12$ ($B>81m^{4}/8$) the effective potential has
always a maximum at $E^{2}$ 
near $r =3m/2$ and near $r =0$. 

In addition, 
for $\lambda \neq 0$ the effective potential has a critical value for the energy, in
which the factor $\sigma ^{2}(r )=0$, for which 
we can determined the
sign of the effective potential: It is positive 
for $E^{2}<\frac{\lambda ^{2}}{\rho _{c}^{2}}(1-\frac{2}{\rho _{c}}+\frac{\alpha }{6\rho _{c}^{4}})$. 
For $\epsilon _{0}$ $\geqslant \epsilon$,
(\textbf{$\rho _{c}=\frac{r_c}{m}$} is the critical value of 
$\rho $ 
%\color{red}
%corresponding the negative energy), 
where the effective potential becomes negative, when $\rho_c$ 
is approached from below)
%\color{black}
the
effective potential is regular and finite, which 
means that the potential
barrier disappears in this case, but the event horizon, which appears there, is not
the effect of minimal length $l$.
%From Fig. \ref{fig7}, the change occurs only at the minimum of 
%the effective potential at $r_e=3m/2$. 

\end{itemize}
\vskip 0.5cm

In conclusion, in all cases the effect of minimal 
length $l$, or maximal
acceleration, 
%\color{red}
are noticeable 
%appears,
%\color{black}
only for small black hole 
masses, as a potential barrier at the horizons 
$0<r_{0}<r_{-}<r_{+}$ in which the effective 
potential diverges or has
singularities. 
%\color{red}
A "small" black hole can still be relatively large,
as $m=10^{18}l$ corresponds to a black hole mass
of about $10^8$kg.
%\color{black}
The height of the barriers increase
with the particle energy.
As a consequence , the formation of a 
larger black hole is stopped. 
%\color{red}
Also the positions of maximal accelerations $r_{\pm 1}$
and $r_{\pm 2}$ were revealed. At these points, the
effective potential has a barrier and a particle
is prevented to pass this barrier.
%\color{black}

\section*{Acknowledgments}

P.O.H. acknowledges financial supoport from PAPIIT-DGAPA
(IN100421).


\begin{thebibliography}{99}

\bibitem{solar} C. M. Will, Living Rev. Relativ. 9 (2006) 3.

\bibitem{einstein1}  A. Einstein, Ann. Math. {\bf 46} 
(1945), 578.

\bibitem{einstein2} A. Einstein, Rev. Mod. Phys. {\bf 20} 
(1948), 35.

\bibitem{born1} M. Born, Proc. Roy. Soc. A {\bf 165} (1938), 291.

\bibitem{born2} M. Born, Rev. Mod. Phys. {\bf 21} (1949), 463.

\bibitem{3} E.R. Caianiello. Lett. Nuovo Cimento {\bf 25} 
(1979), 225.

\bibitem{4} E.R. Caianiello. Lett. Nuovo Cimento {\bf 27} 
(1980), 89.

\bibitem{5} E.R. Caianiello. Il Nuovo Cimento B {\bf 59} 
(1980), 350.

\bibitem{6} E.R. Caianiello, G. Marmo and G. Scarpetta, Il Nuovo Cimento A
{\bf 86} (1985), 337.

\bibitem{7} E.R. Caianiello, La Rivista 
del Nuovo Cimento {\bf 15} (4) (1992)
and references therein.

\bibitem{8} E.R. Caianiello. Lett. Nuovo Cimento {\bf 32} 
(1981), 65.

\bibitem{9} E.R. Caianiello, S. De Filippo, G. Marmo and G. Vilasi. Lett.
Nuovo Cimento {\bf 34} (1982), 112.

\bibitem{21} E.R. Caianiello, S. Capozziello, R. de Ritis, A. Feoli, G.
Scarpetta, Int. J. Mod. Phys. D {\bf 3} (1994), 485.

\bibitem{26} P. O. Hess, W. Greiner, 
Int. J. Mod. Phys. E {\bf 18} (2009), 51.

\bibitem{book} Hess P. O., Sch\"afer M., Greiner W., 
{\it Pseudo-Complex General Relativity}, (Springer, Heidelberg, Germany, 2015). 

\bibitem{PPNP} P. O. Hess, Progr. Part. and Nucl. Phys.,
{\bf 114} (2020), 103809.

%\color{red}
\bibitem{feoli1999} A. Feoli G. Lambiase, G. Papini and G. Scarpetta, Phys. Lett A {\bf 263} (1999), 147.
%\color{black}

\bibitem{caia1999} E. R. Caianiello, M-. Casperini,
G. Scarpetta, Class. Quant. Grav. {\bf 8} (1991), 659.

\bibitem{10} W.R. Wood, G. Papini and Y.Q. Cai, Il Nuovo Cimento B {\bf 104} (1989), 361
and errata ibid (1989), 727.

\bibitem{11} G. Papini, Mathematica Japonica {\bf 41} 
(1995), 81.

\bibitem{12} A. Das, J. Math. Phys. {\bf 21} (1980), 1506.

\bibitem{13} H.E. Brandt, Lett. Nuovo Cimento {\bf 38} 
(1983), 522 and errata
ibid {\bf 39} (1984), 192; Found. Phys. Lett. {\bf 2} 
(1989), 39 and references
therein.

\bibitem{14} M. Toller, Nuovo Cimento B {\bf 102} 
(1988), 261; Int. J. Theor. Phys.
{\bf 29} (1990), 963; Phys. Lett. B {\bf 256} 
(1991), 215.

\bibitem{15} B. Mashoon, Physics Letters A {\bf 143} 
(1990), 176 and references
therein.

\bibitem{16} V.P. Frolov and N. Sanchez, Nucl. Phys. 
B {\bf 349} (1991), 815.

\bibitem{17} A.K. Pati, Europhys. Lett. {\bf 18} 
(1992), 285.

\bibitem{18} A. Feoli, Nucl. Phys. B {\bf 396} (1993), 261.

\bibitem{19} N. Sanchez, in \textquotedblleft Structure: from Physics to
General Systems\textquotedblright\ eds. M. Marinaro and G. Scarpetta (World
Scientific, Singapore, 1993) vol. 1, p. 118.

\bibitem{20} S. Capozziello and A. Feoli, 
Int. J. Mod. Phys. D {\bf 2} (1993), 79.

\bibitem{22} ] A. Crumeyrolle, Ann. de la Fac. des Sciences 
de Toulouse {\bf 4}, s\'{e}rie {\bf 26} (1962), 105.

\bibitem{23} A. Crumeyrolle, Riv. Mat. Univ. Parma {\bf 5} 
(2) (1964), 85.

\bibitem{24} R. L. Clerc, Ann. de L'I.H.P. Section 
A {\bf 12}, No. 4 (1970), 343.

\bibitem{25} R. L. Clerc, Ann. de L'I.H.P. 
Section A {\bf 17}, No. 3 (1972), 227.

\bibitem{kelly} P. F. Kelly and R. B. Mann, Class. and Quant. Grav. {\bf 3} (1986), 705.

\bibitem{27} P. O. Hess, W. Greiner, Int. J.
Mod. Phys. E {\bf 16} (2007), 1643.

\bibitem{28} P. O. Hess, L. Maghlaoui and W. Greiner, 
Int. J. Mod. Phys. E {\bf 19}
(2010), 1315.

\bibitem{visser} M. Visser, Phys. Rev. D {\bf 54} 
(1996), 5116.

\bibitem{Schon2013} T. Schönenbach, G. Gasper, P. O. Hess, T. Boller, A. Müller, M. Schäfer, W. Greiner, Mon. R. Astron. Soc. {\bf 430} (2013), 2999.

\bibitem{Schon2014} T. Schönenbach, G. Gasper, P. O. Hess, T. Boller, A. Müller, M. Schäfer, W. Greiner, Mon. R. Astron. Soc. {\bf 442} (2014), 121.

\bibitem{wave2016} B. P. Abbot, et al., (LIGO Scientific Collaboration and Virgo Collaboration), 
Phys. Rev. Lett. {\bf 116} (2016), 061102.

\bibitem{Nielsen2018} A. Nielsen, O. Birnholz, AN 
{\bf 339} (2018), 298.

\bibitem{Nielsen2019} A. Nielsen, O. Birnholz, AN {\bf 340} 
(2019), 116.

%\bibitem{hessne} Gunther Caspar, Thomas Schonenbach (Frankfurt U., FIAS),
%Peter Otto Hess (Mexico U., ICN), Mirko Schafer, Walter Greiner (Frankfurt
%U., FIAS), Int.J.Mod.Phys. E21 (2012) 1250015.

\bibitem{universe} P. O. Hess and E. L\'opez-Moreno, Universe
{\bf 5} (2019), 191.

\bibitem{hawking} S, Hawking, MNRAS {\bf 152} (1971), 75.

\bibitem{gw} P. O. Hess and E. L\'opez-Moreno, Astr. Nachr.
{\bf 342} (2021), 1034.

%\bibitem{em} E.R. Caianiello, M. Gasperini, G. Scarpetta, II. Nuovo Cimento
%B 105 (1990), p. 259.

\end{thebibliography}
\end{document}